\documentclass[nonacm]{acmart}

\usepackage[most]{tcolorbox}
\usepackage{caption}
\usepackage{array}
\usepackage{diagbox}
\usepackage{hyperref}
\usepackage{supertabular}
\usepackage{booktabs}
\usepackage{float}
\usepackage{enumitem}
\usepackage{graphicx}
\usepackage{comment}
\usepackage{multirow}
\usepackage{tabularx}
\usepackage{colortbl}
\usepackage{longtable}
\usepackage[resetlabels,labeled]{multibib}
\usepackage{subcaption}
\usepackage{makecell}

\acmJournal{CSUR}

\setcopyright{none}
\renewcommand\footnotetextcopyrightpermission[1]{}

\setcopyright{none}
\renewcommand\footnotetextcopyrightpermission[1]{} % removes footnote with conference information in first column
\AtBeginDocument{%
  \providecommand\BibTeX{{%
    \normalfont B\kern-0.5em{\scshape i\kern-0.25em b}\kern-0.8em\TeX}}}

\begin{document}

\title{A Review on C3I Systems' Security: Vulnerabilities, Attacks, and Countermeasures}

\author{Hussain Ahmad}
\email{hussain.ahmad@adelaide.edu.au}
\affiliation{%
 \institution{CREST - The Centre for Research on Engineering Software Technologies, The University of Adelaide, CSCRC - Cyber Security Cooperative Research Centre}
 \country{Australia}}

\author{Isuru Dharmadasa}
\email{isuru.mahaganiarachchige@adelaide.edu.au}

\author{Faheem Ullah}
  \email{faheem.ullah@adelaide.edu.au}

\affiliation{%
 \institution{CREST - The Centre for Research on Engineering Software Technologies, The University of Adelaide}
  \country{Australia}}

\author{M. Ali Babar}
\email{ali.babar@adelaide.edu.au}
\affiliation{%
 \institution{CREST - The Centre for Research on Engineering Software Technologies, The University of Adelaide, CSCRC - Cyber Security Cooperative Research Centre}
 \country{Australia}
}

\authorsaddresses{
Authors' addresses: Hussain Ahmad, hussain.ahmad@adelaide.edu.au; Isuru Dharmadasa, isuru.mahaganiarachchige@adelaide.edu.au; Faheem Ullah, faheem.ullah@adelaide.edu.au; M. Ali Babar, ali.babar@adelaide.edu.au, The University of Adelaide, Australia.}

\renewcommand{\shortauthors}{Ahmad et al.}

\begin{abstract}

Command, Control, Communication, and Intelligence (C3I) systems are increasingly used in critical civil and military domains for achieving information superiority, operational efficacy, and greater situational awareness. Unlike traditional systems facing widespread cyber-attacks, the sensitive nature of C3I tactical operations make their cybersecurity a critical concern. For instance, tampering or intercepting confidential information in military battlefields not only damages C3I operations, but also causes irreversible consequences such as loss of human lives and mission failures. Therefore, C3I systems have become a focal point for cyber adversaries. Moreover, technological advancements and modernization of C3I systems have significantly increased the potential risk of cyber-attacks on C3I systems. Consequently, cyber adversaries leverage highly sophisticated attack vectors to exploit security vulnerabilities in C3I systems. Despite the burgeoning significance of cybersecurity for C3I systems, the existing literature lacks a comprehensive review to systematize the body of knowledge on C3I systems' security. Therefore, in this paper, we have gathered, analyzed, and synthesized the state-of-the-art on the cybersecurity of C3I systems. In particular, this paper has identified security vulnerabilities, attack vectors, and countermeasures/defenses for C3I systems. Furthermore, our survey has enabled us to: (i) propose a taxonomy for security vulnerabilities, attack vectors and countermeasures; (ii) interrelate attack vectors with security vulnerabilities and countermeasures; and (iii) propose future research directions for advancing the state-of-the-art on the cybersecurity of C3I systems.

%Due to the criticality of tactical domains, C3I systems have become a focal point for cyber adversaries. Cyber adversaries exploit security vulnerabilities of C3I systems to perform malicious activities in tactical domains.  Therefore, researchers have proposed several countermeasures for securing the cyberspace of C3I systems. 

%Despite the dire need of securing C3I cyberspace, existing literature lacks a comprehensive review/survey on the cybersecurity of C3I systems.

%However, the sensitive nature of application domains (e.g., battlefield and rescue operations) of C3I systems makes their security a critical concern. For instance, a cyber-attack on military installations can have detrimental impacts on national security. Therefore, in this paper, we review the state-of-the-art on the cybersecurity of C3I systems. In particular, this paper aims to identify the security vulnerabilities, attack vectors, and countermeasures for C3I systems. We gather, analyze and synthesize existing literature available on the cybersecurity of C3I systems. Our survey enabled us to identify 13 security vulnerabilities, 19 attack vectors, and 40 countermeasures for C3I systems. This survey has also revealed several areas for future research and identified key lessons with regards to C3I systems' security. 
\end{abstract}



\maketitle

\section{Introduction} \label{introduction}

\newcolumntype{C}[1]{>{\centering\arraybackslash}m{#1}}

A Command, Control, Communication, and Intelligence (C3I) system is an integration of data gathering sensors, intelligent computing machines, and heterogeneous communication networks for collecting, storing, analyzing, and transmitting information in tactical domains under the supervision of authorized commanders \cite{shenoy1987command}. Due to the novel intelligence (e.g., Artificial Intelligence (AI) techniques \cite{niu2021development}) and cognitive agility \cite{natter2010review}, C3I systems enable organizations to attain and sustain information superiority, operational efficacy, increased situational awareness, real-time decision support, swift communication, and enhanced collaboration among heterogeneous C3I assets during their operations \cite{mineo2004advanced}. Moreover, C3I command systems ensure strict compliance with organizational chain-of-commands, which prevents a C3I asset from violating an informed course of actions in tactical operations. Therefore, C3I systems are increasingly used in sensitive civil and military domains, such as search-and-rescue missions, healthcare, transportation, fireguard, battlefield, airfield, and many other applications \cite{djordjevitowards}, where timely data transmission and plan execution are of prime concern. For example, the UK government leveraged C3I systems to effectively respond to healthcare emergencies resulting from COVID-19 \cite{hutchings2021command}.

%C3I systems offer an unconventional assistance in data collecting, storing, processing, and transmitting among heterogeneous systems in a dynamic tactical environment [2].

%These unique characteristics of   One recent example is the use of C3I system in handling the COVID-19 pandemic in UK, where C3I was successfully used to manage the healthcare crises and load as a result of spike in COVID cases [6].

\begin{figure*}[!tbp]
  \centering
  \captionsetup{justification=centering}
  \includegraphics[trim=0 165 0 0,clip, scale = 0.4]{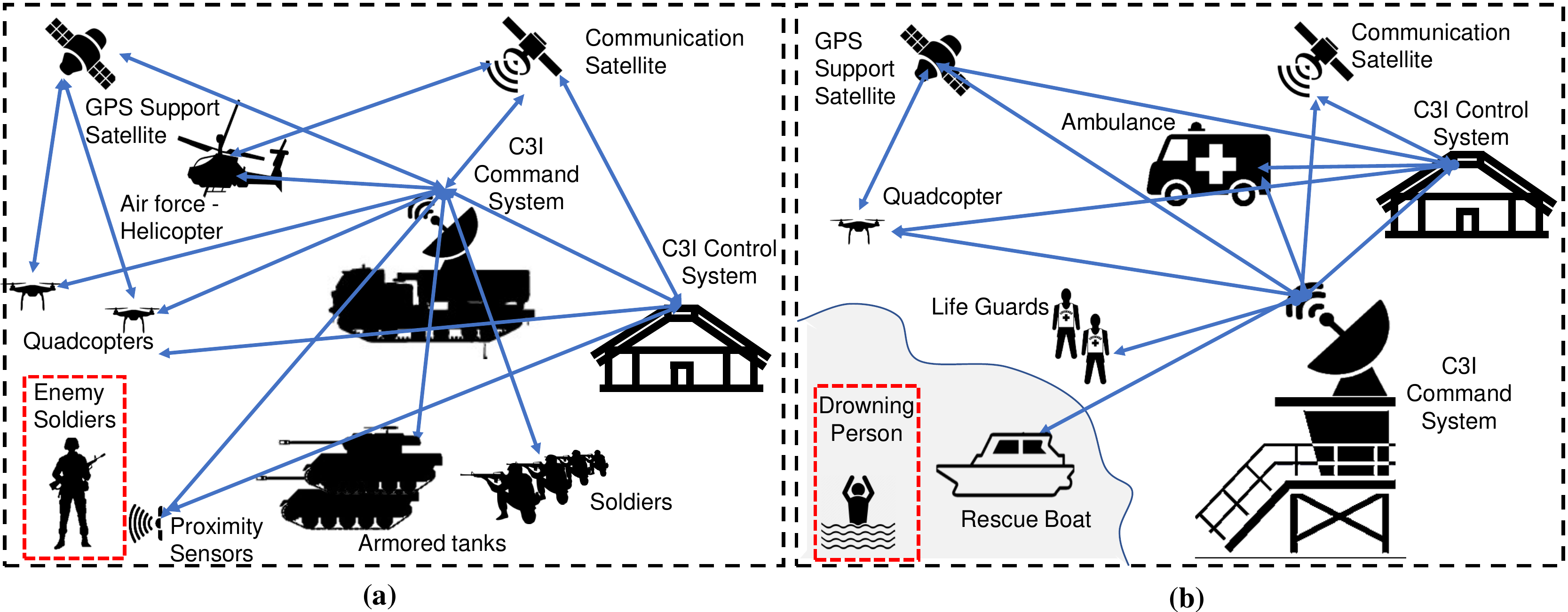}
  \caption{Abstract view of a C3I system application in (a) military operation and (b) rescue mission.}
  \label{fig:applicationscenario} 
\end{figure*}

To illustrate the significance of C3I systems in tactical operations, Figure~\ref{fig:applicationscenario} shows two application scenarios of a C3I system for a  military and a civil domain - (a) battlefield and (b) rescue mission. In the battlefield scenario, both C3I command and C3I control systems separately collect tactical data (e.g., positions and activities of enemy soldiers) through different sensing devices such as quadcopters and proximity sensors. The control system processes the raw data with the help of a C3I intelligence unit and provides classified information to the command system. Accordingly, the C3I command system executes a required action plan by coordinating multiple military troops, such as soldiers, helicopters, and armored tanks, for achieving mission objectives. C3I communication systems, such as satellite links and other technologies (e.g., 4G/5G and RF links), enable mobile C3I assets (e.g., helicopters and quadcopters) to effectively collaborate during tactical operations. Using a similar approach, but with a different goal, a C3I command system directs lifeguards and rescue boats to save a drowning person in the illustrated rescue mission scenario. In both scenarios, C3I systems achieve information superiority, operational efficacy, and situational awareness by producing required intelligence through collecting and processing sensitive data during tactical operations. 

As illustrated, C3I systems are used in sensitive domains such as military missions and search-and-rescue scenarios. By sensitive domains, we mean domains where the consequences of actions going wrong are quite detrimental. For instance, feeding incorrect target location to an aircraft for shelling during a military operation can result in severe undesired damage including loss of human lives and damage to infrastructure. The sensitive nature of C3I application domains makes their cybersecurity a critical concern. For example, considering the military operation presented in Figure~\ref{fig:applicationscenario}.(a), if proximity sensors get compromised due to an adversarial attack, a C3I system collects falsified data and generates erroneous commands accordingly, resulting in mission failures. Historical events also indicate that cybersecurity breaches of C3I systems led to major military defeats. For example, the compromise of Enigma machines (part of C3I communication systems) was one of the main reasons behind the loss of Germans in world war II \cite{ormrod2014coordination}. Similarly, the Afghan war documents leak hosted in C3I infrastructure, published by WikiLeaks in 2010, is one of the biggest data breaches in military history \cite{malik2012application}. The unauthorized disclosure comprised approximately 91,000 US confidential military records regarding the war in Afghanistan. Another cybersecurity violation was reported by the Washington Post\footnote{https://www.washingtonpost.com/world/national-security/with-trumps-approval-pentagon-launched-cyber-strikes-against-iran/2019/06/22/250d3740-950d-11e9-b570-6416efdc0803\_story.html} in 2019 when the US officially launched cyber-attacks on Iranian C3I military instalments. The cyber-attacks impaired C3I systems controlling missile and rocket launchers. Concerning the C3I civil domain applications, ransomware attacks on the Los Angeles hospital\footnote{https://www.theguardian.com/technology/2016/feb/17/los-angeles-hospital-hacked-ransom-bitcoin-hollywood-presbyterian-medical-center} and the San Francisco public transit\footnote{https://www.theguardian.com/technology/2016/nov/28/passengers-free-ride-san-francisco-muni-ransomeware} not only disrupted their C3I operations, but also caused unauthorized disclosure of sensitive information. Hence, a large number of cyber-attacks \cite{attackDoD} targeting C3I systems have become a grave concern for the cybersecurity of C3I systems.

%figure 01

Given the increased complexity of tactical operations (e.g., DoD C3I modernization strategy \cite{c3modern}), the contemporary C3I systems have started leveraging advanced features of modern technologies, e.g., blockchain \cite{akter2019highly} and cloud computing \cite{koo2020security}), to meet the stringent operational requirements such as rapid response, reliability and service assurance in tactical environments. The incorporation of the state-of-the-art technologies in the contemporary C3I systems have increased the potential risk of sophisticated cyber-attacks such as Advanced Persistent Threats (APTs). The cyber threat surfaces can be present in any C3I system component such as databases, web servers, and communication networks. A cyber-attack on a C3I system is considered successful when an adversary exploits system's vulnerabilities resulting in adverse consequences such as unauthorized disclosure, tampering, and unavailability of sensitive information along with the monetary and reputation loss \cite{gansler2004information}. Furthermore, the use of the state-of-the-art technologies for executing cyber-attacks exacerbates the adverse impacts on C3I systems \cite{alghamdi2011proposed}.

The increased impact of cyber-attacks emphasizes the necessity of designing, developing, and adopting adequate security measures to secure C3I systems. Therefore, security experts, system designers and developers employ defensive strategies, which are broadly known as countermeasures, to secure C3I systems from cyber-attacks. For example, countries like the United States, the United Kingdom, and organizations like North Atlantic Treaty Organization (NATO) have proposed architectural frameworks such as NATO Architecture Framework (NAF), Ministry of Defense architecture framework (MoDAF), and The Department of Defense Architecture Framework (DoDAF) to enhance the cybersecurity of C3I related tactical systems \cite{alghamdi2011proposed}. Furthermore, developments such as designing the state-of-the-art technical services for C3I domains \cite{MacBCyber}, the adaptation of emerging technologies like big data, IoT, 5G communication \cite{forecast2025} and establishment of cyber defense systems \cite{iranmilitary} are also noticeable in the present context concerning the cybersecurity of C3I systems. In addition to countermeasures, it is also important to identify the common vulnerabilities that are often exploited by cyber adversaries. Equally important is the identification of the attack vectors that are used by adversaries to exploit the vulnerabilities. Such exploration of vulnerabilities and attack vectors facilitate researchers and security experts to develop the required security safeguards/countermeasures for securing C3I systems. 

\begin{figure*} [!tbp]
  \centering
   \includegraphics[trim=0 880 100 0, clip, scale = 0.3]{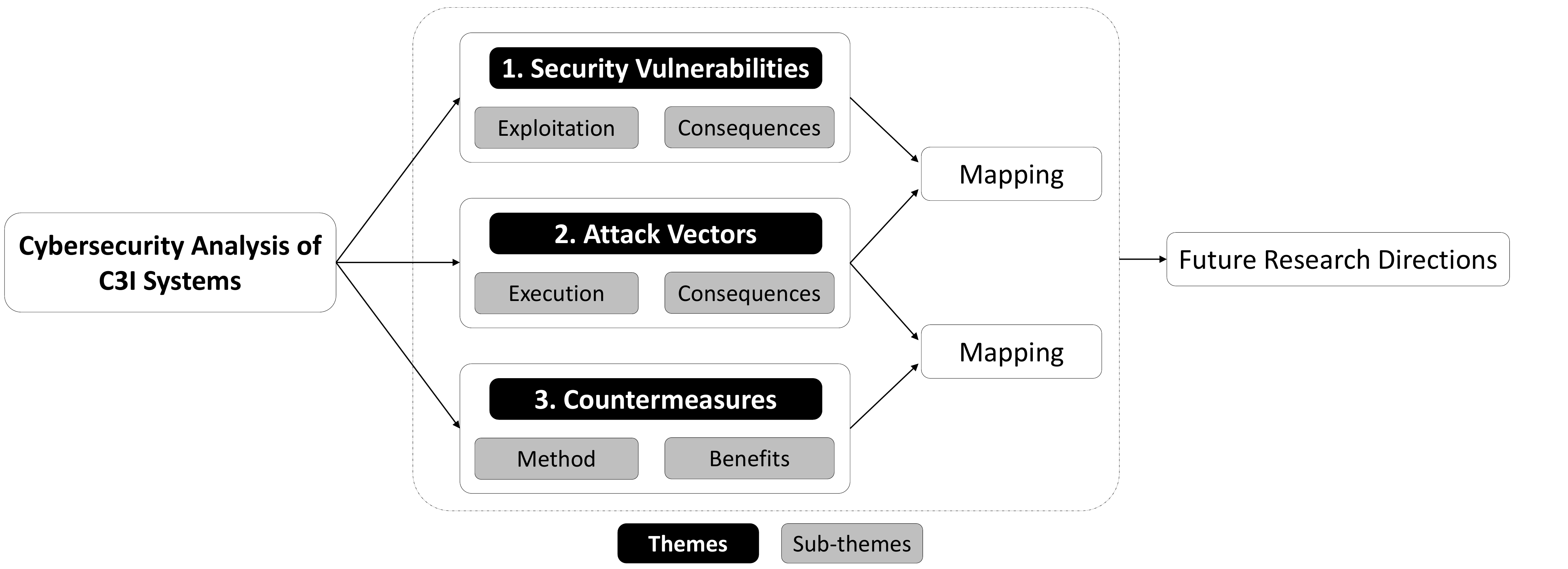}
  \caption{Identified themes for investigating cybersecurity of C3I systems}
  \label{fig:themes} 
\end{figure*}

Although researchers have proposed several countermeasures and identified vulnerabilities and attack vectors for C3I systems, to the best of our knowledge, there is no survey/review study aimed at reviewing the existing literature for systematizing a body of knowledge on the cybersecurity of C3I systems. To fill this gap, our research has systematically gathered, analyzed and synthesized the state-of-the-art on the cybersecurity of C3I systems. For reviewing the literature on this topic, we have considered only the peer-reviewed studies published after 2000 in order to provide the most contemporary and validated insights on the cybersecurity of C3I systems. Based on our analysis of the data extracted from the reviewed studies, we have divided this paper into three themes (i.e., security vulnerabilities, attack vectors and countermeasures) as presented in Figure~\ref{fig:themes}. For an in-depth cybersecurity analysis of C3I systems, we divide each theme into two sub-themes. For example, we describe exploitation details and their detrimental consequences for each security vulnerability identified in the literature. Similarly, we report the execution of attack vectors and their adverse impacts on C3I systems. The analysis and reporting of these themes have led us to identify relationship among them and the future research areas for practitioners and researchers.

%Table~\ref{tab:related-work} presents a comparison between the focused areas of the existing reviews/surveys and our survey, which shows that our survey is the only review study on the cybersecurity of C3I systems.
%For conducting this study, we started this survey by gathering primary studies related to the cybersecurity of C3I systems. In addition to the academic search engines (e.g., Google Scholar), we mainly consulted the seven well-known digital libraries presented in Table~\ref{tab:datasources}.
%For in depth security analysis, we divide each theme into two sub-themes. For example, we explore vulnerabilities exploitation details and their detrimental consequences in C3I systems while investigating the C3I systems' security vulnerabilities through the existing literature. Similarly, we report cyber-attacks execution details and their impacts on C3I systems for each attack vector.
%presented in Table~\ref{tab:datasources}.

\noindent \underline{\textbf{Our Contributions:}} In summary, our survey makes the following contributions.

\begin{itemize}
    \item It presents a comprehensive analysis of C3I systems’ security vulnerabilities identified in the literature. Each security vulnerability has been meticulously described along with its exploitation details and consequences of exploitation on C3I systems. The identified security vulnerabilities are innovatively categorized based on C3I system components. 
    %\item First comprehensive survey on the cybersecurity of C3I systems.
    \item It provides a high-level investigation of the attack vectors applicable to C3I systems. Each attack vector has been critically examined focusing on its execution methodologies and adverse impacts on C3I systems. The attack vectors are also categorized based on C3I system components i.e., command, control, communication, and intelligence. 
    %\item Identification and categorization of 13 security vulnerabilities, 19 attack vectors and 40 countermeasures reported in the existing literature.
    \item It gives an overarching analysis of the countermeasures reported in the literature for securing the cyberspace of C3I systems. The methods and benefits offered by each countermeasure have been described. The identified countermeasures are categorized based on the development and operational phases of C3I systems. Furthermore, each category has been critically investigated providing its benefits and limitations. 
    %\item Develop a relationship between security vulnerabilities, attack vectors and countermeasures for securing C3I systems.
    %\item Identification of critical factors for C3I security and providing recommendations to practitioners for securing C3I systems
    \item It presents a combined analysis of the security vulnerabilities, attack vectors and countermeasures for C3I systems. The exhaustive analysis not only results in developing a distinctive relationship of attack vectors with security vulnerabilities and countermeasures, but also identifying future research areas for advancing the state-of-the-art on the cybersecurity of C3I systems. 
    %\item Identification of areas for advancing the state-of-the-art on the cybersecurity of C3I systems.  
\end{itemize}

It is important to mention that we analyzed the cybersecurity of C3I systems and their derivative systems in this review. These derivative systems include Command, Control, Communication, Computers, and Intelligence (C4I) Systems; Command, Control, Communication, Computers, Cyber, and Intelligence (C5I) systems; Command, Control, Communication, Computers, Intelligence, surveillance, and reconnaissance (C4ISR); and many other combinations (e.g., C5ISR and C6ISR) \cite{russell2018internet}. However, for the ease of readers, we use the term \textit{C3I systems} to refer to the C3I and its all other derivative systems throughout the paper.

\noindent \underline{\textbf{Paper Structure: }} The remainder of this survey is organized as follows. Section \ref{overview} reports an overview of C3I systems. Section \ref{vulnerabilites}, Section \ref{attacks} and Section \ref{countermeasures} describe the security vulnerabilities, attack vectors, and countermeasures, respectively. Section \ref{discussion} presents an analysis of our research findings, which mainly includes the mappings among the research findings and the identification of future research directions. Lastly, Section \ref{conclusion} concludes the survey.

\section{An Overview of C3I systems}  \label{overview}

\begin{figure*} [!bp]
  \centering
   \includegraphics[trim=0 10 350 0, clip, scale = 0.39]{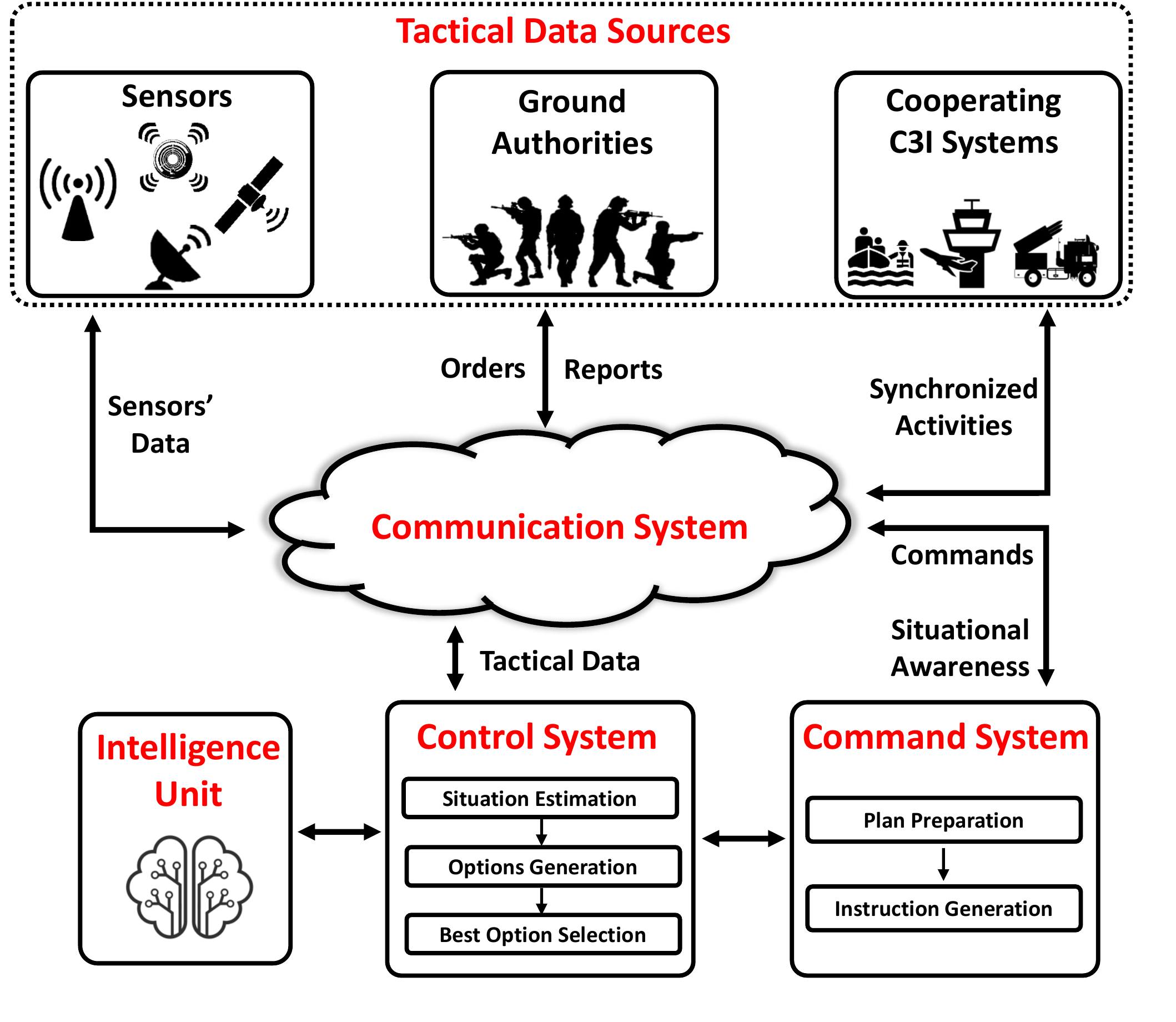}
  \caption{Role of each C3I system's component in tactical operations}
  \label{fig:c3ioverview} 
\end{figure*}

This section provides an overview of the C3I system to help understand the findings reported in the subsequent sections. In particular, we describe C3I system components and their functionalities in tactical operations. A C3I system is mainly composed of four components:  command system, control system, communication network, and intelligence unit. These C3I components operate in conjunction with each other to execute critical civil and military operations. At the start of a tactical operation, both C3I command and control systems separately collect tactical information through different data sources (e.g., sensors, field commanders and C3I systems operating in tactical environments) \cite{djordjevitowards}, as depicted in Figure~\ref{fig:c3ioverview}.

\textit{\textbf{Control System:}} C3I control systems process received tactical information to generate an action plan required for achieving C3I mission objectives. For this purpose, C3I control systems employ data computing machines (e.g., single board computers \cite{matthews2018harnessing}), data controllers (e.g., PLCs and SCADA \cite{fisher2012hacms}), and storage devices (e.g., solid-state drives \cite{spanjer2008security}). The raw data collected from tactical domains are mainly processed through the following three steps \cite{djordjevitowards}. \textit{Step 1:} A tactical situation is estimated by extracting its relevant features from the received information. \textit{Step 2:} The estimated situation is compared with a desired mission’s outcome to generate possible action plans. \textit {Step 3:} An optimal action plan is selected from possible options according to the availability and requirement of strategic resources (e.g., Quality of Service, cost and power). Finally, the C3I control system shares the optimal action plan with the corresponding C3I command system for its validation and implementation. It is important to note that a C3I intelligence unit facilitates control system activities, which is described later in this section.

\textit{\textbf{Command System:}} Tactical data sources provide situational awareness (e.g., geographic location and movement of tactical assets) to C3I commanders through C3I command systems such as C3I web interfaces \cite{salmon2007measuring}, ecological interfaces \cite{bennett2008ecological} and android applications \cite{inforsuperiority}). C3I commanders first analyze both the shared situational awareness and optimal action plan of C3I control systems to prepare an effective strategy against adversaries for achieving a set goal. Then, they implement a finalized action plan by giving instructions to field commanders and autonomous systems operating in tactical environments through command interfaces. C3I command interfaces ensure a strict chain-of-commands in tactical operations, which prevents unauthorized use of confidential information by cyber adversaries. 

% (e.g., and two-way radio transceiver \cite{kouwen2018digital}). 

\textit{\textbf{Communication System:}} C3I communication system supports all the inter-and intra-communication between C3I components and tactical data sources during collection, processing, and dissemination of data in tactical environments. For this purpose, C3I communication systems connect widely distributed, mobile and heterogeneous C3I assets (e.g., sensors, autonomous C3I systems, and ground authorities) for data transmission and overall communication. To integrate heterogeneous C3I assets, C3I communication networks consist of versatile and diversified data transmission links that include terrestrial line-of-sight (e.g., Link-16 and very high frequency), terrestrial beyond-line-of-sight (e.g., joint tactical radio systems and cooperative engagement capability data distribution systems) and satellite (e.g., narrowband and wideband) communication facilities \cite{malik2012application}. Moreover, other data transmission protocols and technologies such as MANET \cite{bednarczyk2015performance}, RF links \cite {balzano2007high}, 4G/5G \cite {oh2021plan}, and SATCOM \cite{attila2018technologies} are also used in C3I networks. For data translation between heterogeneous assets, C3I communication systems employ efficient data exchange models (e.g., JC3IEDM \cite{de2015approach} and MIEM \cite{hayes2008maritime}) to enhance interoperability and coalition among heterogeneous tactical assets during C3I operations. 

\textit{\textbf{Intelligence Unit:}} A C3I intelligence unit assists C3I systems in data processing and effective decision-making \cite{russell2018internet}. Different AI tools, such as Machine Learning (ML) and Deep Learning (DL) models, are used to introduce intelligence in C3I operations \cite{schubert2018artificial}. For example, ML-based dimensionality reduction algorithms, reported in \cite{khalid2014survey}, are used to extract the relevant features from the raw information received from C3I data sources to estimate an emerging tactical situation. Similarly, AI-based optimization approaches, reported in \cite{swarnkar2020artificial}, are used for selecting an optimal action plan. C3I intelligence capabilities not only improve situational awareness and operational efficacy \cite{wignessefficient}, but also provide cognitive agility to C3I commanders for effective decision-making \cite{voke2019artificial}.

%ML algorithms such as Principal Component Analysis (PCA) \cite{gewers2021principal} and linear discriminant analysis \cite{xanthopoulos2013linear} 

% Section 03: Research Methodology - End

% Section 04: Vulnerabilities - Start

\section{Security Vulnerabilities}  \label{vulnerabilites}
\begin{figure*} [!hbp]
  \centering
\includegraphics[trim=0 30 0 0, clip, scale = 0.45]{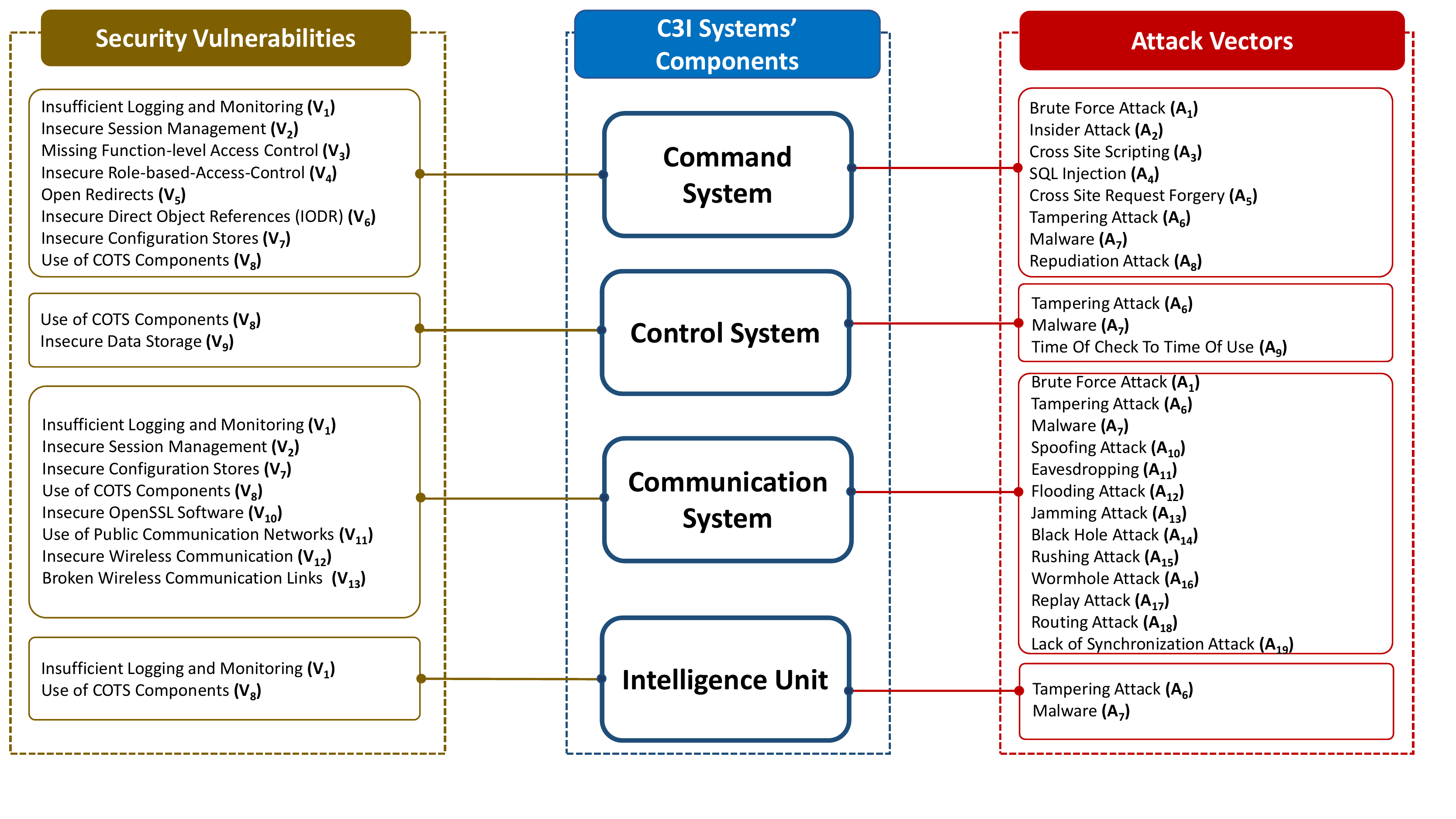}
  \caption{Security vulnerabilities and attack vectors associated with command, control, communication, and intelligence components of a C3I system}
  \label{fig:vul-attacks} 
\end{figure*}

\begin{figure*} [!htbp]
  \centering
\includegraphics[trim=0 180 255 0, clip, scale = 0.5]{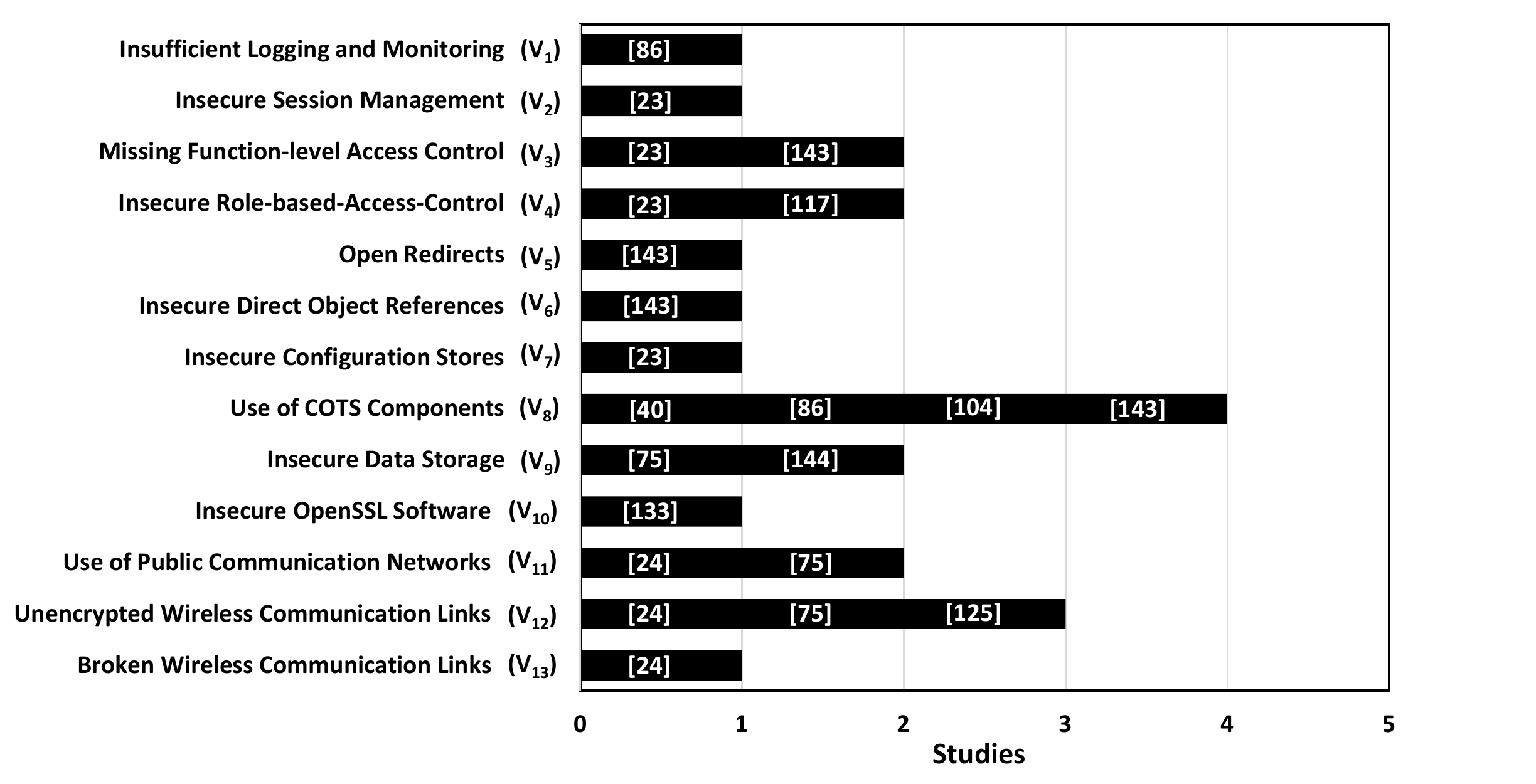}
  \caption{Identified C3I system security vulnerabilities and their sources/references}
  \label{fig:vul-source} 
\end{figure*}

This section reports the findings related to the theme 1, security vulnerabilities of C3I systems. In general, a security vulnerability refers to any weakness, fault, issue, or loophole in a system that an attacker can exploit to harm a system or its' users. Based on this definition, we have identified 13 security vulnerabilities denoted those as V\textsubscript{1}, V\textsubscript{2}, V\textsubscript{3}, ... V\textsubscript{13}. We categorized the identified vulnerabilities based on their associated C3I system components (i.e., Command, Control, Communication and Intelligence). Since some vulnerabilities (e.g., V\textsubscript{1} and V\textsubscript{8}) relate to multiple C3I components, we have explained the variants of these vulnerabilities under each component. Furthermore, for each vulnerability, we have provided examples from Common Vulnerabilities and Exposures (CVE)\footnote{https://cve.mitre.org/} database considering their relevance and applicability to C3I systems. Figures \ref{fig:vul-attacks} and \ref{fig:vul-source} illustrate the overall categorization of security vulnerabilities and attack vectors, and vulnerabilities with their extracted studies, respectively.

\subsection{Vulnerabilities in C3I Command System} \label{vul-command}

\textit{Insufficient Logging and Monitoring (V\textsubscript{1}):} C3I command systems should be equipped with continuous monitoring methods such as intrusion detection \citep{liao2013intrusion} to enable real-time unauthorized access identification and prevention. However, system architectures such as Service Oriented Architecture (SOA) do not consist of in-built intrusion detection and continuous monitoring mechanisms since SOA-based systems are typically used in enterprise applications \cite{jormakka2009intruder}. Hence, SOA-based C3I systems can recognize neither intruders in real-time nor compromised systems in the long run. Furthermore, considering the mechanisms such as, OpenPegasus Common Information Model (CIM) servers that can be used to monitor C3I systems' hardware performance and health, it can contain this security vulnerability as OpenPegasus 2.7 CIM does not record the failed login attempts (CVE-2008-4315). By exploiting this vulnerability, adversaries can execute numerous attacks to unauthorizedly access a C3I system without being detected or notified to the relevant authorities.
%https://owasp.org/www-project-top-ten/2017/A10_2017-Insufficient_Logging%2526Monitoring

\textit{Insecure Session Management (V\textsubscript{2}):} Many web-based C3I command systems use sessions to store user information, including session ID in the server-side storage, to uniquely identify each user and their authorized operations \cite{gutzmann2001access}. The failure to design user authentication mechanisms with security measures such as cryptographically strong session IDs \cite{alghamdi2010enhancing} and secure session termination policies \cite{vlsaggio2010session} can enable attackers to retrieve session data from the server and unauthorizedly access C3I command systems to break or affect the strict chain-of-commands followed in C3I systems \cite{rezakhani2018novel}. For example, when C3I systems employ NoSQL database software such as Infinispan, the C3I system can get affected from this vulnerability since Infinispan-9.4.14 lacks proper session fixation protection (CVE-2019-10158).  
%https://hdivsecurity.com/owasp-broken-authentication-and-session-management

\textit{Missing Function-level Access Control (V\textsubscript{3}):} Similar to any other system, C3I command systems also have multiple users with different access rights. For example, system administrators have higher privileges than ordinary users (e.g., data operators), which they leverage through dedicated administrative interfaces. When systems are not designed correctly to provide function-level access control (i.e., strict user authorization for each function), users with low access rights can elevate their privileges and execute unsanctioned operations \cite{alghamdi2010enhancing, rezakhani2018novel, missingFunction, missingFunction2}. For instance, when C3I systems leverage service management software, such as IBM Jazz, users can access and delete restricted data and resources from C3I systems by exploiting this vulnerability (CVE-2019-4194).

%https://blog.detectify.com/2016/07/13/owasp-top-10-missing-function-level-access-control-7/

\textit{Insecure Role-Based Access Control (V\textsubscript{4}):} As explained in V\textsubscript{3}, C3I command systems require to manage users with multiple roles and access privileges. Role-Based Access Control (RBAC) is a widely used access control method, where each user has a defined user role(s) with a set of permitted actions \cite{ferraiolo2003role}. While RBAC systems can facilitate role-based access control over C3I command system functionalities and data, misconfigured RBAC systems (e.g., inaccurate mapping between user roles and their associated functions) can make mission-critical data be manipulated and erased by attackers \cite{alghamdi2010enhancing, maschino2003access}. For example, Oracle Solaries 11.1 operating system powered C3I systems are affected by this vulnerability, allowing local users to conduct operations that are restrict to their user roles (CVE-2013-5875). 
%https://owasp.org/www-project-proactive-controls/v3/en/c7-enforce-access-controls

\textit{Open Redirects (V\textsubscript{5}):} In web-based C3I command systems, users rely on Uniform Resource Locators (URL) to navigate between internal and external C3I systems. Therefore, it is critical to ensure that these web links are validated and directed only to secure C3I systems and domains. For example, when C3I system users utilize web-based Cisco Webex meetings to conduct online meetings, remote attackers can send malicious URLs and steal C3I system user credentials by redirecting them into insecure web pages since this software does not strictly validate user given URLs (CVE-2021-1310) \cite{wichers2013owasp}. In that case, an attacker can steal the credentials of a C3I admin and use the fraudulently received credentials to impersonate and execute attacks on C3I systems \cite{rezakhani2018novel, unvalidatedlinks2}.  
%https://blog.detectify.com/2016/08/15/owasp-top-10-unvalidated-redirects-and-forwards-10/

\textit{Insecure Direct Object References (IODR) (V\textsubscript{6}):} Similar to V\textsubscript{5}, IODR vulnerability is also associated with the insecure access control of web-based C3I command systems. In web development, it is common to use the name or key of an object to populate web pages dynamically. Therefore, if C3I command systems do not validate user requests to access data residing in C3I systems, an adversary can manipulate legitimate requests by injecting restricted internal object references to escalate their privileges, horizontally and vertically \cite{kumarshrestha2015identification, insecureDirect2, rezakhani2018novel}. For example, when C3I systems use LogonBox Nervepoint Access Manager for user authentication and identity management, attackers can exploit IODR vulnerability and unauthorizedly retrieve C3I system user details (CVE-2019-6716). 
%https://blog.detectify.com/2016/05/25/owasp-top-10-insecure-direct-object-reference-4/

\textit{Insecure Configuration Stores (V\textsubscript{7}):} A typical C3I command system consists of multiple servers configured for various functionalities such as, web services, email servers and file transfer services. The configuration store of a server manages all the setting details of the implemented security methods, approaches, and techniques \cite{configStore}. Server misconfigurations and disclosure of access details to unauthorized third parties can result in an insecure configuration store \cite{alghamdi2010enhancing}. An attacker who gains access to a command system configuration store can disable the implemented security mechanisms and make the C3I system vulnerable to numerous cyber-attacks. For example, when C3I systems are equipped with application management software such as ManageEngine Applications Manager, malicious authenticated users can escalate their privileges by capitalizing on this vulnerability. As a result, these adversaries can take complete control of the overall C3I system (CVE-2019-19475). 
%https://owasp.org/www-pdf-archive/OWASP_Top_10_2007.pdf (A8 is closely related)

\textit{Use of COTS Components (V\textsubscript{8}):} Many C3I systems utilize Commercial off-the-shelf (COTS) components to reduce the cost and time spent on the development. However, some of these software components may contain security loopholes due to inadequate testing and review procedures \cite{biagini2016modelling}. Even with the tested software components, some are not patched for known vulnerabilities \cite{rezakhani2018novel}. It is important to note that when C3I systems adopt these third-party software associated with open-source codebases, attackers can exploit vulnerabilities by launching attacks on C3I systems \cite{jormakka2009intruder, littlejohn2017mission}. For example, when C3I authentication systems employ vulnerable third-party JSON Web Token Libraries (CVE-2021-41106) \cite{BibEntry2021Nov}, C3I command systems face the threat of unauthorized access to mission-critical data. 

\subsection{Vulnerabilities in C3I Control System}

\textit{Use of COTS Components (V\textsubscript{8}):} Similar to Section \ref{vul-command} - V\textsubscript{8}, C3I control systems can become vulnerable to cyber-attacks due to the use of compromised third-party software for tasks such as security-critical data processing, storing, monitoring and visualizing. Therefore, when C3I control systems are equipped with software from malicious vendors, without conducting adequate security testing to speed-up the development process, these systems can contain back-doors and pre-installed malware, secretively \cite{Nist2021Nov}. For example, when C3I systems are equipped with IBM InfoSphere servers for storing mission-critical data, attackers can steal these data by exploiting these servers' inherited insecure third-party domain access vulnerability (CVE-2021-29875).

\textit{Insecure Data Storage (V\textsubscript{9}):} C3I control systems store diverse types of data, including security critical information, to expedite C3I operations through increased situational awareness and tactical decision support. Thus, the security of data at rest is essential in C3I systems. Data loss can happen in C3I systems due to many reasons, such as hardware and software failures, user negligence, and adversary attacks \cite{insecureData, insecureData2, gkioulos2017security, romero2013cealician}. For example, when C3I system data are stored in Couchbase Servers, attackers can access these data since these Couchbase servers store security-critical data in plain text (CVE-2021-42763). 
%https://owasp.org/www-project-mobile-top-10/2016-risks/m2-insecure-data-storage

\subsection{Vulnerabilities in C3I Communication System}

\textit{Insufficient Logging and Monitoring (V\textsubscript{1}):} Similar to Section \ref{vul-command} - V\textsubscript{1}, C3I communication systems also should be equipped with continuous logging and monitoring mechanisms to detect malicious traffic coming from compromised C3I nodes. For example, when C3I communication systems leverage network devices equipped with JUNOS operating system, which lacks proper resource allocation and monitoring methods, attackers can send malicious traffic to these devices and make them unavailable for legitimate C3I data communications (CVE-2021-31368).  

\textit{Insecure Session Management (V\textsubscript{2}):} C3I communication systems facilitate efficient and secure data communication among widespread C3I nodes. For example, battlefield data from UAVs must be transmitted to the C3I control unit for further processing and orders from the command system must be communicated with military troops to take necessary actions. The session layer in the Open Systems Interconnection (OSI) model is responsible for creating, synchronizing and terminating communication channels between devices \cite{alani2014guide}. Thus, security methods like SSL/TLS certificates \cite{singh2015secure} are required in the session layer to enable secure communication between legitimate C3I nodes. However, the fallacious implementation of these security measures allows adversaries to make C3I communication networks unavailable for critical data communication leading to mission failures (CVE-2021-40117).  

\textit{Insecure Configuration Stores (V\textsubscript{7}):} Similar to Section \ref{vul-command} - V\textsubscript{7}, C3I systems have to employ security configuration management systems such as Cisco Firepower Management Center (FMC) for implementing and maintaining C3I communication systems' security mechanisms. However, as FMC system stores user data in plain text, authenticated local attackers can retrieve these details and disable the implemented security defenses of C3I communication networks by impersonating as system administrators (CVE-2021-1126).  

\textit{Use of COTS Components (V\textsubscript{8}):} Similar to Section \ref{vul-command} - V\textsubscript{8}, C3I communication systems can become vulnerable due to the use of compromised COTS devices (e.g., routers and switches) and network management tools (e.g., Wireshark and SolarWinds). For example, due to a vulnerability in SolarWinds, a network monitoring and management software, nine US federal agencies have been compromised, and attackers have gained access to data and emails of these systems (CVE-2021-35212) \cite{willett2021lessons}. Similarly, any C3I system that uses SolarWinds third-party software possesses the threat of unauthorized access to C3I mission-critical data. 

% \textit{Insecure Software Development Practices (V\textsubscript{9}):} Similar to Section 3.1 - V9, C3I communication networks can become vulnerable when designed and developed not adhering to computer network security standards. For example, network perimeter security mechanisms such as firewalls must be implemented and configured correctly in C3I communication networks to prevent intruders from accessing C3I systems and malicious data packets being sent into edge networks \cite{wang2009computer}. Attackers can access C3I systems and their data and vandalize communication through DOS attacks when these security measures are absent in C3I networks  \cite{wang2009computer}, \cite{malik2012application}. 

\textit{Insecure OpenSSL Software (V\textsubscript{10}):} OpenSSL is an encryption software that is used in C3I web systems to support secure communication among C3I web systems and users \cite{Lee2015May}. It ensures an active receiver is available at the other end of the communication link through "heartbeat" messages. Since OpenSSL software had not implemented a strict "heartbeat" message validation mechanism, attackers can exploit these messages to access server Random Access Memory (RAM) \cite{o2015cyber, Lee2015May, Synopsys2020Jun}. Therefore, attackers can exploit this vulnerability to unauthorizedly retrieve C3I mission-critical information and other security related data like credentials, from the C3I system server RAM (CVE-2014-0160). 

\textit{Use of Public Communication Networks (V\textsubscript{11}):} Some C3I systems use public networks such as the internet for data communication to minimize cost \cite{alghamdi2011proposed}. Since public networks lack strict security measures (e.g., secure Virtual Private Networks (VPN) \cite{segura2003secure}) by default, transferred data is susceptible to adversarial attacks relating to unauthorized access and data manipulation \cite{and2021somebody, alghamdi2011proposed, gkioulos2017security}. For example, when C3I components communicate through public networks without secure VPN channels, an attacker can view and tamper transmitted data leading to compromised C3I operations. However, even the use of VPN channels does not guarantee the confidentiality and integrity of the data transmitted through public communication channels as some VPN software are not designed with adequate security mechanisms (e.g., insecure input validation - CVE-2021-1519).  

\textit{Unencrypted Wireless Communication Links (V\textsubscript{12}):} C3I systems depend on wireless communication networks mainly due to the cost and impracticality of using wired technologies with mobilized units \cite{zahmati2011transmission}. Compared to wired communication technologies, wireless links are susceptible to an increased number of attacks since transceivers have limited control over the propagation range and direction of wireless signals \cite{zou2016survey, alghamdi2011proposed, mursia2011simulation, gkioulos2017security}. Like V\textsubscript{3}, when C3I wireless networks are not designed with necessary security measures like physical layer encryption \cite{tang2021encrypted}, adversaries can execute numerous attacks to disable the communication links between C3I systems and decrease the overall situational awareness of C3I operations. For example, when C3I communication systems leverage wireless devices with internal data encryption process errors (e.g., Broadcom WiFi client devices - CVE-2019-15126), C3I mission-critical data can be decrypted by attackers leading to data confidentiality breaches.    
%https://owasp.org/www-pdf-archive//OWASP_Mumbai_2008.pdf

\textit{Broken Wireless Communication Links (V\textsubscript{13}):} As described in V\textsubscript{12}, C3I systems highly depend on wireless communication technologies to assure connectivity among geographically widespread C3I assets. However, wireless communication links frequently get affected due to excessive movements of C3I assets in hostile conditions \cite{alghamdi2011proposed}. For example, when a legitimate node gets disconnected due to broken wireless links, attackers can penetrate the C3I system by joining as the disconnected legitimate node (CVE-2020-24586). As a result, the attacker gets access to the C3I system amidst the legitimate node becomes vulnerable due to the lack of communication and decreased situational awareness.

\subsection{Vulnerabilities in C3I Intelligence Unit}

\textit{Insufficient Logging and Monitoring (V\textsubscript{1}):} While Section \ref{vul-command} explains the vulnerability V\textsubscript{1} in terms of lack of intrusion detection and continuous monitoring mechanisms in C3I command systems, here, we focus on two main aspects related to the application and security of AI/ML-based methods for monitoring and logging in C3I systems. Firstly, we emphasize the lack of AI/ML-based methods employed by C3I systems for intrusion detection methods. For example, machine learning classifiers such as K-nearest neighbor (K-NN), Support Vector Machine (SVM), and Artificial Neural Networks (ANN) have been widely used in anomaly detection leveraging security event logs \cite{tsai2009intrusion, kumar2017practical}. Hence, the C3I intelligence unit can incorporate such mechanisms to detect intruders and subsequently implement preventive methods to stop attackers' access to C3I systems. The failure to incorporate these intelligent intrusion detection methods will allow intruders to access and execute malicious attacks on C3I systems by compromising systems' integrity. Secondly, we highlight the necessity of continuously monitoring AI/ML models' performance (e.g., accuracy) in C3I intelligence units. AI/ML models suffer from data and model drifts leading to inaccurate inferences \cite{andresini2021insomnia}. For example, when C3I systems employ AI/ML models for intrusion detection that are trained for other domains, the accuracy of these models can be significantly lower due to data drift. Therefore, the AI/ML models employed in C3I systems must be continuously monitored, tested and verified to ensure that these models provide accurate results.     

\textit{Use of COTS Components (V\textsubscript{7}):} Similar to Section \ref{vul-command} - V7, C3I intelligence units can become vulnerable to cyber-attacks when these units employ insecure third-party AI/ML models and frameworks to generate knowledge (e.g., situational awareness) from raw data (e.g., sensor data). For example, tactical C3I systems' can employ the TensorFlow Deep Learning (DL) framework with Convolutional Neural Networks (CNN) to identify enemy territories from satellite images. However, the TensorFlow framework possesses a Denial of Service (DoS) threat when used with the NumPy package, which is commonly leveraged in AI/ML model development (CVE-2017-12852). Therefore, when these insecure AI/ML packages are used in C3I intelligence units, attackers can exploit this vulnerability to make the intelligence unit unavailable or non-responsive, impacting C3I operations and leading to mission failures.

% \textit{Insecure Software Development Practices (V\textsubscript{8}):} Similar to Section 3.1 - V8, C3I intelligence units can become vulnerable to adversarial attacks when AI/ML models employed in these systems have not followed secure development practices. For example, AI/ML models must be tested against adversarial machine learning attacks to ensure the robustness and accuracy of these models \cite{huang2011adversarial} to be employed with mission-critical data such as surveillance videos and satellite images. Adversaries can execute attacks such as model inversion, model inference, and data poisoning on insecure AI/ML models \cite{chakraborty2018adversarial} to generate false intelligence and mislead C3I operations when such security practices are not followed.

% Section 05: Attacks - Start

\section{Attack Vectors} \label{attacks}
\begin{figure*} [!tbp]
  \centering
\includegraphics[trim=0 30 175 0, clip, scale = 0.5]{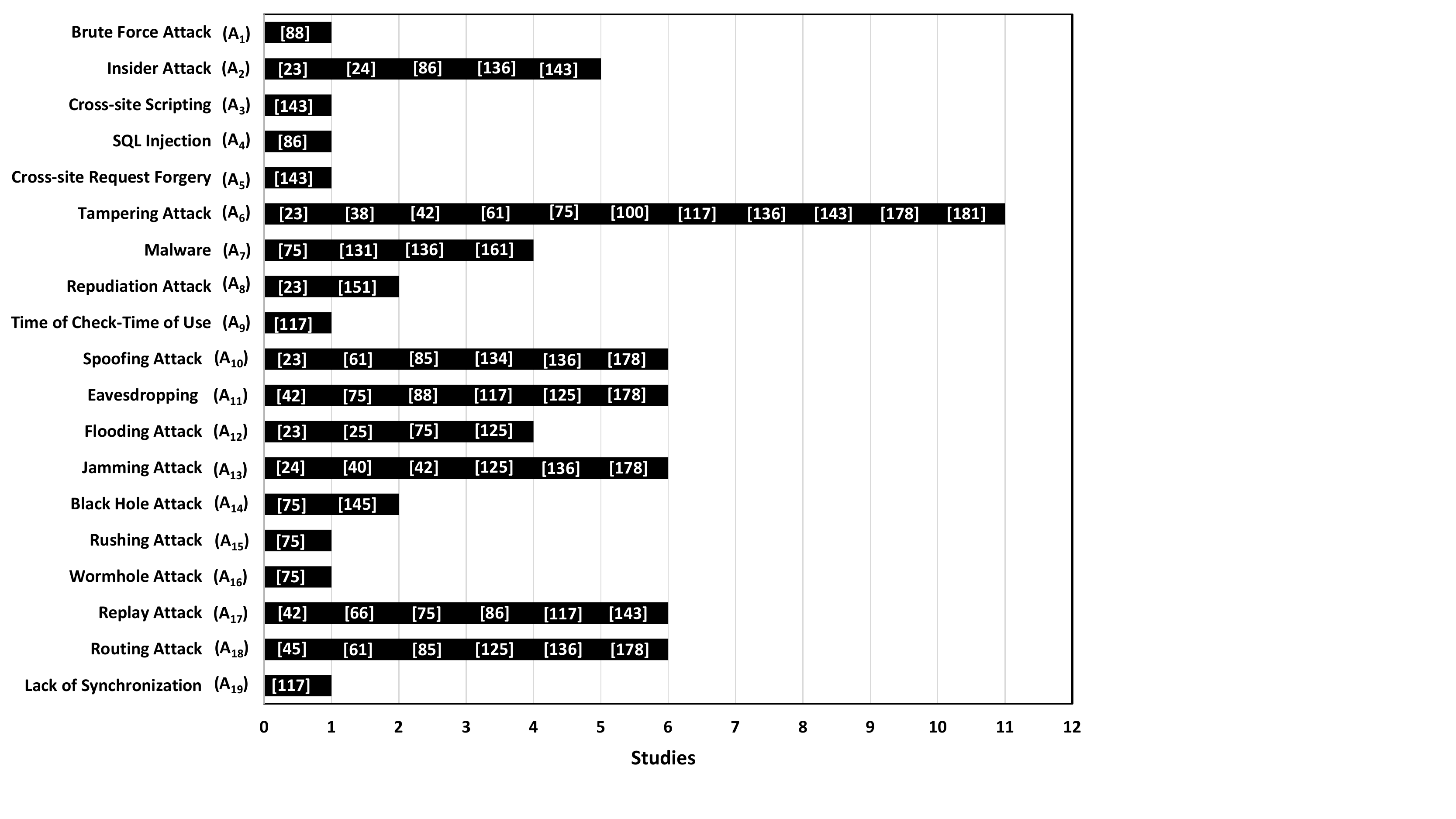}
  \caption{Identified C3I system attack vectors and their sources/references}
  \label{fig:attack-source} 
\end{figure*}

This section reports the findings related to the theme 2, attack vectors against C3I systems. An attack vector refers to a methodology used by an attacker to exploit a vulnerability in C3I systems. We have identified 19 attack vectors targeting the cybersecurity of C3I systems through the existing literature (Figure~\ref{fig:attack-source}). We denote the 19 attack vectors by A\textsubscript{1}, A\textsubscript{2}, A\textsubscript{3}, …., A\textsubscript{19} to facilitate their referencing throughout the paper. Similar to C3I systems vulnerabilities (Section \ref{vulnerabilites}), we categorize the attack vectors based on their applicability on C3I system components, as presented in Figure~\ref{fig:vul-attacks}. Since the tampering attack (A\textsubscript{6}) and the malware (A\textsubscript{7}) can be executed on any C3I system component, we report both the attack vectors for each C3I component. In the following, we describe the attack vectors with their execution (sub-theme 1) and impact (sub-theme 2) details for corresponding C3I system components.

\subsection{Attack Vectors for C3I Command System} \label{attack-command}
%In this section, we report the attack vectors that target C3I command interfaces to disturb command operations in tactical domains.
\textit{Brute Force Attack (A\textsubscript{1}):} A brute force attack is used to get unauthorized access to C3I command interfaces \cite{kang2006towards}. Hackers use different intrusion mechanisms, such as trial-and-error methods \cite{dave2013brute} and session IDs exploitation \cite {endler2001brute}, to obtain secret information (e.g., cryptographic keys and login credentials) of C3I commanders. As a result, intruders not only receive real-time situational awareness, but also perform malicious activities (e.g., generating fraudulent instructions) through compromised C3I command interfaces.

\textit{Insider Attack (A\textsubscript{2}):} Insider refers to a malicious C3I system's operator who has legitimate access to C3I command systems. An insider attack is executed when insiders deliberately or mistakenly neglect security protocols while using C3I command interfaces \cite{alghamdi2010enhancing, perkinson2012lessons, jormakka2009intruder, alghamdi2011proposed, rezakhani2018novel}. Such actions result in leakage of sensitive information and termination of tactical operations. Due to the legitimate access privileges of insiders, detecting or preventing insider attacks is cumbersome in C3I command systems \cite{alghamdi2011proposed}.

\textit{Cross-Site Scripting (A\textsubscript{3}):} Cross-Site Scripting (XSS) is an attack vector, used against web-based C3I command interfaces, in which a malicious script is injected into the output of a C3I command interface \cite{rezakhani2018novel}. When a C3I commander accesses a compromised interface, an installed malicious script is activated, which enables intruders to steal sensitive information (e.g., users activities) from C3I command systems \cite{vogt2007cross}. Session hijacking and user impersonation are the results of XSS \cite{vogt2007cross}.

\textit{SQL Injection (A\textsubscript{4}):} Structured Query Language (SQL) is a code that is used to get sensitive information from C3I storage facilities through web-based C3I command interfaces. SQL injection occurs when intruders inject a malicious SQL query to access a C3I database \cite{jormakka2009intruder, maschino2003access}. Consequently, intruders exfiltrate, destroy, or manipulate sensitive information stored in C3I databases. For example, Romanian hackers execute a SQL injection attack against the US army website in 2010. As reported by DARKReading\footnote{https://www.darkreading.com/risk/u-s-army-website-hacked}, the hackers successfully gained access to 75 databases containing sensitive information army personnel.

\textit{Cross-Site Request Forgery (A\textsubscript{5}):} Cyber adversaries create fraudulent HTTP links to perform malicious activities through C3I command interfaces \cite{rezakhani2018novel}. If an authenticated C3I commander clicks on deceptive links, corresponding malicious actions are executed. Since vulnerable web interfaces cannot distinguish between a legitimate request and a forged one sent by an authorized user \cite{jovanovic2006preventing}, it is difficult for such command interfaces to detect forged requests. A cross-site request forgery enables attackers to execute malicious commands in tactical domains through C3I command interfaces.

\textit{Tampering Attack (A\textsubscript{6}):} A tampering attack, reported in \cite{alghamdi2010enhancing, fang2015research, perkinson2012lessons, kwon2019identification, zachary2003decentralized, maschino2003access, brade2006need, rezakhani2018novel, gkioulos2017security, bernier2012metrics, zehetner2004information}, is executed when intruders perform malicious activities to manipulate tactical information communicated through C3I command interfaces. For example, web parameter tampering attacks \cite{dalai2012novel} tamper users credentials, operational commands and information communicated through C3I web interfaces by using POST requests. Consequently, C3I commanders cannot implement a required action plan, which may lead to the failure of a C3I mission.

\textit{Malware (A\textsubscript{7}):} Malware are malicious software \cite{theron2019autonomous, ormrod2014coordination, perkinson2012lessons, gkioulos2017security} that are used to penetrate a C3I system for executing malicious activities such as unauthorized access, data modification and exfiltration. Malware are usually spread through malicious emails \cite{qbeitah2018dynamic}, drive-by downloading \cite{cova2010detection}, and external removable devices \cite{poeplau2012honeypot} in C3I command systems. As a result, malware restrict authorized users to access C3I interfaces, provide hackers remote access to C3I interfaces, and steal sensitive information about command operations.  

\textit{Repudiation Attack (A\textsubscript{8}):} A repudiation attack occurs when intruders modify the records of activities executed in C3I command operations \cite{alghamdi2010enhancing, shaneman2007enhancing}. By using different malicious tactics (e.g., log injection attack \cite{datta2020cyber}), attackers change the stored information of actions taken by C3I commanders or destroy logs of their own malicious activities during a C3I operation. Consequently, corrupt log files make the validity of command actions sceptical and create confusion in C3I chain-of-commands.

\subsection{Attack Vectors for C3I Control System}
%This section describes the attack vectors for data processing machines and storage facilities of C3I control systems.
\textit{Tampering Attack (A\textsubscript{6}):} A tampering attack, described in Section \ref{attack-command}, is also used to modify and fabricate tactical information in data computing and storage systems of C3I control units. Intruders perform malicious activities, such as corrupting database configurations and injecting malicious code in program executable, to manipulate sensitive information of C3I control systems \cite {peddoju2020file}. Consequently, C3I control systems incorrectly estimate tactical situations and generate falsified action plans for command systems.

\textit{Malware (A\textsubscript{7}):} Similar to the C3I command system (Section \ref{attack-command}), malware are also harmful to C3I control systems. The use of third-party software \cite{Nist2021Nov}, insecure communication connections \cite{damshenas2013survey}, and vulnerable operating systems \cite{damshenas2013survey} are common ways to inject malware in data computing machines and storage devices. As a result, malware cause data manipulation, exfiltration and disruption in control system activities, which distorts an action plan generation process required to achieve C3I tactical objectives. For example, Stuxnet malware disrupted the operations of computer-aided control systems (e.g., SCADA) used in Iranian nuclear power plant along with other 30,000 IP addresses \cite{malik2012application}.

\textit{Time Of Check To Time Of Use (A\textsubscript{9}):} C3I control systems check the availability of strategic resources (e.g., operational equipment and human resource) before devising an action plan to achieve mission objectives. A time of check to time of use attack \cite{maschino2003access} is executed when intruders perform malicious activities (e.g., malicious code injection and resource consumption) on the available resources in between the time of their check and usage to invalidate the result of a check operation \cite{pu2006methodical}. Consequently, C3I control systems prepare an action plan with unavailable/compromised resources, which executes unintended actions in C3I operations.

\subsection{Attack Vectors for C3I Communication System}

%In this section, we describe the attack vectors targeting the data transmission links of C3I communication networks.
\textit{Spoofing Attack (A\textsubscript{10}):} Cyber adversaries impersonate a legitimate C3I node to get connected with C3I networks for executing malicious activities (e.g., stealing tactical information and inserting malware). Different attack vectors, such as IP spoofing \cite{ehrenkranz2009state}, ARP spoofing \cite {trabelsi2009arp}, DNS spoofing \cite{abdelmajid2020location} and MAC spoofing \cite {madani2020mac}, are used to steal the identity of a legitimate C3I node through different communication layers. For example, open-source software, such as Kismet \cite{kismet} and Ethereal \cite{ethereal}, are used to get or change a valid MAC address of a C3I system \cite{vaidya2016review}. Another form of spoofing attack is GPS spoofing \cite{perkinson2012lessons} in which attackers generate counterfeit GPS signals by using commercial off-the-shelf products \cite{chapman2017gps} to provide falsified position, navigation and timing information to C3I nodes in tactical environments.

%For this purpose, an attacker steals the identity of legitimate C3I nodes by using spoofed IP, ARP, DNS or MAC addresses \cite{babu2010comprehensive}. Open-source software like Kismet \cite{kismet} or Ethereal \cite{ethereal} are used to get or change a valid MAC address of a C3I system \cite{vaidya2016review}. Similarly, spoofed email addresses and websites can also be used for executing the spoofing attack to get sensitive information from a C3I network. The spoofing attack damages the confidentiality of data \cite{alghamdi2010enhancing}.

%\textit{Spoofing Attack (A\textsubscript{5}):} Domain Name Server (DNS) spoofing attack alters the DNS records stored in a DNS cache of C3I command interfaces. The malicious DNS entries returns a fraudulent response that directs the

%\textit{GPS Spoofing (A\textsubscript{21}):} C3I systems operating in a tactical environment are heavily dependent on GPS signals to determine their location in time and space. Accurate GPS locations of C3I nodes are also imperative for decision-makers to design collaborative tasks for a common objective. As GPS signal structure is in the public domain, an attacker can easily generate low frequency, counterfeit GPS signals by using a commercial off-the-shelf equipment \cite{chapman2017gps}. Spoofed GPS signals provide false position coordinates to a targeted C3I node \cite{perkinson2012lessons}. As a result, an attacked C3I node performs its assigned activities according to a spoofed position and time; hence, deteriorates the node authentication and overall C3I activity \cite{malhi2020security}.

\textit{Eavesdropping (A\textsubscript{11}):} Eavesdropping, also known as a man-in-the-middle attack, is a passive attack vector by which attackers secretly listen to the communication between two C3I nodes \cite{zachary2003decentralized, kang2006towards, maschino2003access, mursia2011simulation, brade2006need, gkioulos2017security}. Open source network monitoring and packet sniffing tools, such as \textit{Wireshark} and \textit{Tcpdump} \cite{goyal2017comparative}, are used to eavesdrop on C3I communication links. Eavesdropping attack results in an unauthorized disclosure of tactical information. \cite{malhi2020security}. For example, Britishers eavesdropped on German military communications through compromised Enigma machines during World War II, which was one of the main reasons behind the defeat of German military \cite{lycett2011breaking}.  

\textit{Flooding Attack (A\textsubscript{12}):} A flooding attack, reported in \cite{alghamdi2010enhancing, mursia2011simulation, gkioulos2017security, alghamdi2010generic}, is executed when an attacker sends large traffic volumes towards a target C3I node to disrupt its services for other C3I nodes in a tactical environment. Large traffic volumes, such as SYN flooding \cite{yuan2008lab} and PING flooding \cite{kumar2006ping}, consume available bandwidth of a server near to a targeted C3I node. Therefore, legitimate data packets coming from other C3I nodes cannot be transmitted from the affected server. Consequently, a target C3I node gets disconnected from a C3I communication network.

\textit{Jamming Attack (A\textsubscript{13}):} Jamming attacks make a C3I system unavailable for other C3I nodes in a C3I communication network \cite{perkinson2012lessons, zachary2003decentralized, alghamdi2011proposed, mursia2011simulation, brade2006need, biagini2016modelling}. Hackers employ different jamming strategies, such as constant jammer and deceptive jammer, to disable data transmission links associated to a target C3I node in a C3I communication system  \cite{vadlamani2016jamming}. As a result, services of an affected C3I system, connected with the disabled channels, become unavailable for other C3I tactical nodes.

\textit{Black Hole Attack (A\textsubscript{14}):} In tactical MANETs, a route discovery process to find a shortest route for data transmission is necessary and inevitable \cite{tseng2011survey}. Hackers exploit this MANET feature by sending a fake route reply message, that a malicious node has a required shortest path, during a route discovery process \cite{gkioulos2017security, roopa2018intelligent}. Consequently, a source C3I node establishes a data transmission route to a destination C3I node through the malicious node, which leads to communication interruption and sensitive information exfiltration.

\textit{Rushing Attack (A\textsubscript{15}):} A rushing attack \cite{gkioulos2017security} is executed when an attacker node receives a route request (RREQ) during a route discovery process in tactical MANETs. The attacker node immediately sends the RREQ packet to all C3I nodes before any other C3I node forwards the same RREQ packet in the network \cite{goyal2017attacks}. In this way, all other C3I nodes consider the RREQ coming from a legitimate C3I node as duplicate, so they repudiate the legitimate RREQ. As a result, the attacker node is always included in a data transmission route, which may cause denial-of-service and eavesdropping (\textit{A\textsubscript{11}}) in tactical MANETs.

\textit{Wormhole Attack (A\textsubscript{16}):} A wormhole attack \cite{gkioulos2017security} occurs when at least two attacker nodes locate themselves at strategic positions in a tactical MANET. The attacker nodes tunnel RREQ packets between each other during a route discovery process. When a destination node receives the RREQ packets transmitted through the tunnel, the destination node finds the malicious route a shortest route in the network and discards all other RREQ packets received from other legitimate C3I nodes \cite{datta2012security}. In this way, attacker nodes become part of a data transmission route in the C3I tactical MANET. A wormhole attack may lead to data tampering (\textit{A\textsubscript{6}}), man-in-the-middle attack (\textit{A\textsubscript{11}}), and data exfiltration in C3I domains.

\textit{Replay Attack (A\textsubscript{17}):} A replay attack, reported in \cite{maschino2003access, jormakka2009intruder, brade2006need, rezakhani2018novel, gkioulos2017security, furtak2016security}, is carried out in three steps. \textit{Step 1:} Monitor C3I communication links by using network monitoring tools \cite{svoboda2015network}. \textit{Step 2:} Intercept sensitive information such as login credentials and C3I action plan details. \textit{Step 3:} Replay the intercepted data packets to deceive a receiver C3I node. Consequently, legitimate C3I nodes consider attacker nodes as authentic C3I nodes, which leads to unauthorized disclosure of sensitive information in C3I systems.

\textit{Routing Attack (A\textsubscript{18}):} Routing attacks are executed on routing protocols of C3I communication networks to disrupt C3I services in tactical environments \cite{fang2015research, jinfeng2017effects, perkinson2012lessons, zachary2003decentralized, mursia2011simulation, chudzikiewicz2019procedure}. Cyber adversaries use different malicious tactics such as routing table overflow \cite{xiao2011based} and routing table poisoning \cite{tayal2013survey} to execute routing attacks on C3I communication systems. As a result, C3I systems cannot make connections with other legitimate C3I nodes, which makes their services unavailable in C3I networks.

\textit{Lack of Synchronization Attack (A\textsubscript{19}):} Time synchronization among widely distributed C3I systems is ensured by using different protocols (e.g., reference broadcast synchronization \cite{elson2002fine}, timing-sync Protocol \cite{ganeriwal2003timing}, and flooding time synchronization \cite{maroti2004flooding}) to execute collaborative C3I operations. However, these time synchronization protocols are not securely designed \cite{manzo2005time}. Therefore, hackers compromise these protocols to share falsified timing information in a distributed C3I environment, which creates a lack of time synchronization among C3I nodes \cite{maschino2003access}. Consequently, asynchronous C3I nodes adversely disturb the operational coordination in time-critical C3I operations.

\textit{Brute Force Attack (A\textsubscript{1}):} Network authentication mechanisms, such as Kerberos \cite{neuman2005kerberos} and WPA/WPA2 \cite{adnan2015comparative}, are used to protect C3I communication systems in tactical environments. These authentication mechanisms require users credentials (i.e., username and password) for allowing authorized users to access network resources. Cyber adversaries use state-of-the-art brute force attack tools (e.g., Reaver\footnote{https://www.kali.org/tools/reaver} and Fern-Wifi-Cracker\footnote{https://www.kali.org/tools/fern-wifi-cracker}) for guessing authorized users credentials to get connected with C3I networks. As a result, adversaries can eavesdrop and intercept sensitive information transmitted through C3I communication links.

\textit{Tampering Attack (A\textsubscript{6}):} Intruders modify and fabricate sensitive information transmitted through C3I communication links. For this purpose, attackers first penetrate a C3I network by using different attack mechanisms (e.g., black hole attack (A\textsubscript{14}) and spoofing attack (A\textsubscript{10}), then, they either modify in-transit data packets or inject their own malicious data into the transmitting sensitive data. As a result, tampered information distorts situational awareness, C3I operational activities and decision-making processes.

\textit{Malware (A\textsubscript{7}):} Insecure communication links facilitate malware propagation from one system to another system in a C3I tactical environment \cite{damshenas2013survey}. In this way, hackers create a network of compromised C3I systems, called botnet, to perform malicious activities in a C3I network. For example, botnet executes distributed denial of service attack, reported in \cite{bingman2016c4isr}, by conducting the flooding attack (A\textsubscript{12}) on a target C3I system to disrupt its C3I communication services. Consequently, an affected C3I system cannot communicate with other legitimate systems in a C3I network, which damages C3I tactical operations.

% Moreover, Brand et al. \cite{brand2011threat} have introduced a concept of malware rebirthing bonet to prevent malware detection by an anti-virus software. According to their proposed idea, malware rebirthing suite takes already inserted malwares as an input and generates a new version of malware with different functionalities to avoid its detection. As a result, decision-makers or legitimate users loss their confidence on available data; a hacker can take control of a C3I system, and makes it unavailable \cite{brand2011threat}.
% Firstly, an attacker hijacks many C3I units distributed in a tactical environment by inserting malware in them, then, each compromised system executes flooding attack on a targeted C3I system \cite{brand2011threat}. Consequently, the targeted C3I system cannot communicate with other legitimate systems in a C3I network.

\subsection{Attack Vectors for C3I Intelligence Unit}
%This section reports the attack vectors, reported in the existing literature, against the AI tools used in C3I systems.
\textit{Tampering Attack (A\textsubscript{6}):} In addition to other C3I components, the tampering attack is also executed on C3I intelligence units. For example, data poisoning attacks \cite{steinhardt2017certified}, a type of adversarial attacks \cite {ren2020adversarial}, add malicious samples into the training datasets of ML/DL models to manipulate C3I intelligence operations during data processing. As a result, infected ML/DL models misinterpret a C3I tactical situation and prepare a falsified action plan accordingly, which eventually damages the decision-making process of C3I command systems.

\textit{Malware (A\textsubscript{7}):} To detect malware in C3I systems, security experts train ML/DL models (e.g., MalConv \cite{raff2018malware}) with expected malware samples. However, advanced adversaries make changes in malware, such as changing header fields and instruction sequence, to evade ML/DL models for malware detection in C3I systems. These evasive variants of malware are known as \textit {adversarial malware binaries} \cite{kolosnjaji2018adversarial} that successfully penetrate C3I systems to perform malicious activities, described in Section \ref{attack-command}, without being detected by ML/DL models.

%For example, Kolosnjaji et al. \cite{kolosnjaji2018adversarial} have introduced a gradient-based attack vector for generating evasive variants of malware samples, also known as ,  to avoid their detection by ML/DL models. Consequently, malware successfully penetrate C3I systems to perform malicious activities as described in Section.

% Section 05: Attacks - End
% Section 06: Defenses - Start

%figure 12
\begin{figure*}[!bp]
  \centering
  \includegraphics[trim=230 560 0 0,clip, scale = 0.38]{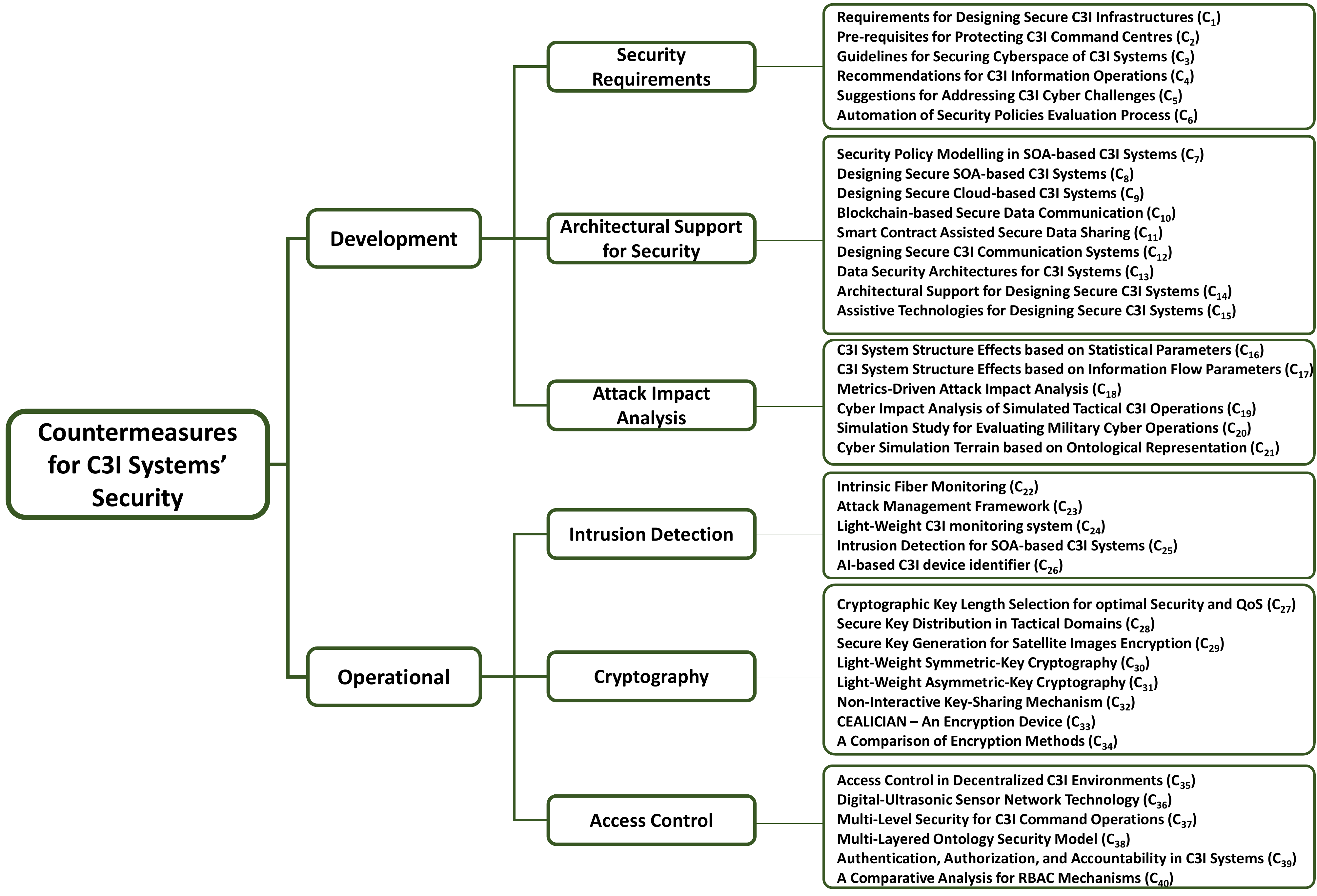}
  \caption{Identified countermeasures and their respective categories.}
  \label{fig:countermeasures} 
\end{figure*}

\section{Countermeasures}  \label{countermeasures}

This section reports the findings related to the theme 3, countermeasures proposed for securing C3I systems. Countermeasures refer to the safeguards or defenses used for protecting C3I systems. We have extracted 40 countermeasures from the reviewed papers. As opposed to the security vulnerabilities and attack vectors, countermeasures are not directly aligned with C3I system components. Therefore, we categorize the extracted countermeasures into two broad categories: development and operational. The countermeasures in the development category are the ones that leverage techniques (e.g., security requirements analysis and security design patterns/tactics) during the development process of C3I systems to help construct a secure C3I system. Contrary to this, the countermeasures in the operational category aim to secure a C3I system while the system is in operation. In Figure~\ref{fig:countermeasures}, C\textsubscript{1}, C\textsubscript{2}, and so on specifies the identifier of the countermeasure used throughout the paper to refer to the respective countermeasures.
%Then, we further categorize the respective countermeasures based on the proposed mechanisms (e.g., cryptography and access control) for the secure development and operations of C3I systems.%Fig.~\ref{fig:countermeasures} shows the categorization of the countermeasures. 

\subsection{Development}

As described in Section \ref{overview}, C3I systems collect, process, store and transmit critical data to support tactical operations in hostile environments. Therefore, the security of a C3I system must be considered as an important quality attribute while developing C3I systems. In other words, security must not be left as an after-thought rather it needs to be considered right from the start of the development process \cite{alghamdi2010enhancing, alghamdi2011proposed}. %Secure C3I system development refers to a process of defining, designing and evaluating the security of C3I systems.%
As depicted in Figure~\ref{fig:countermeasures}, we categorized the countermeasures in the development category into the following three categories. Moreover, we report the benefits and limitations of each category.
%The existing literature has reported security requirements, secure design patterns and attack impact analysis for developing secure C3I systems. 

\subsubsection{Security Requirement Analysis}

Before designing/implementing a C3I system, it is important to analyze, specify, and understand the security requirements of a C3I system keeping in view its operating environment. Such analysis and understanding help the designers of a C3I system to incorporate corresponding security measures for addressing the specified security requirements. Therefore, researchers have reported security requirements (\textit{C\textsubscript{1} to C\textsubscript{6}}) for developing secure C3I systems.

%Security requirement analysis of C3I systems helps in understanding existing security challenges and guidelines for designing, implementing, and operating C3I systems.

%It is always fine to conduct a detailed requirement analysis before designing a C3I system. This pre-analysis helps security experts in understanding existing security challenges and requirements for designing, implementing and operating C3I systems. Similarly, it is also imperative to investigate the security violations in C3I systems after a successful execution of attack vectors. This post-analysis signifies the severity of attack vectors and enables developers to develop secure C3I systems

%\subsubsection{Proactive Investigation}
%As discussed, proactive investigation report the security requirements for developing C3I systems. In the following, we describe the proactive investigative countermeasures, \textit{ C\textsubscript{31} to C\textsubscript{36}}, for designing a secure C3I system.

OZTURK et al. \cite{ozturk2019challenges} (\textit{C\textsubscript{1}}) have recommended the use of digital certificates, digital signatures, firewalls, and smartcards for tactical information security. Moreover, secure design patterns, such as \cite{dangler2013categorization} and \cite{dougherty2009secure}, are suggested for developing secure C3I systems. Similarly, Li et al. \cite{li2011challenges} (\textit{C\textsubscript{2}}) have suggested the use of spread spectrum, private lines, instant/directional radio frequency communication and frequency hopping approaches for avoiding interception of communication signals between different C3I components and tactical data sources (e.g., sensors). The study \cite{li2011challenges} has also recommended the use of camouflage techniques and equipment (e.g., anti-radar and anti-infrared devices) especially in command systems to prevent enemies' reconnaissance. Besides, Bingman \cite{bingman2016c4isr} (\textit{C\textsubscript{3}}) has presented several guidelines focused on safeguarding critical information in contested C3I cyberspace environments. The proposed guidelines address the challenges including but not limited to information prioritization, risk assessment, secure infrastructure, and commercial networks. The authors in \cite{zehetner2004information} (\textit{C\textsubscript{4}}) describe security engineering and information security principles ( e.g., ISO/IEC 17799 and AS/NZS 4360) for data security in C3I systems. For Canadian C3I cyber operations, Bernier et al. \cite{bernier2010understanding} have reported various suggestions (\textit{C\textsubscript{5}}) to address the challenges of dynamic nature and indistinct boundaries of cyber environments. For example, the authors have suggested the integration of computer network attack, defense and exploitation operations with each other during C3I operations to get a holistic picture of their cyberspace. Concerning the challenges of NATO federated mission networks, Lopes et al. \cite{lopes2021cyber} (\textit{C\textsubscript{6}}) have investigated the use of three technologies (i.e., software-defined networking, network security function, and network function virtualization) for automating the security policy evaluation and implementing secure information exchange functions between different components of a C3I system.  

\begin{tcolorbox}[left=0pt, top=0pt, right=0pt, bottom=0pt, boxrule=0.75pt]
\textbf{\textit{Benefits of Security Requirement Analysis}}
\begin{itemize}[leftmargin=*]
    \item \small \textit{{Upfront consideration of securing C3I systems}}
    \item \textit{Guides to identify and adapt secure technologies such as communication
    protocols, encryption mechanisms}
    \item \textit{Adaptable to different types of C3I domains (e.g., military and civil)}
\end{itemize}
\end{tcolorbox}
\begin{tcolorbox}[left=0pt, top=0pt, right=0pt, bottom=0pt, boxrule=0.75pt]
\textbf{\textit{Limitations of Security Requirement Analysis}}
\begin{itemize}[leftmargin=*]
    \item \small \textit{{Security requirement specification and analysis is costly in terms of time and budget}}
    \item \textit{Further customizations are required when adapting the generic security guidelines to design systems’ security}
\end{itemize}
\end{tcolorbox}

\subsubsection{Architectural Support for Security} \label{security_by_design}

In this section, we describe the countermeasures, \textit{C\textsubscript{7} to C\textsubscript{15}}, focused on securing C3I systems at the architectural/design level. The reviewed studies have used different types of architectural styles for designing C3I systems. These styles include service-oriented architecture, cloud-based architecture, blockchain-based architecture, and security architectures designed explicitly for C3I systems. 
%We have identified a few commonly used system architectures (e.g., service-oriented architecture and cloud-computing architecture) and novel C3I system oriented architectures (e.g., autonomous cyber defense) proposed to develop secure C3I systems. Therefore, we categorized the selected countermeasures under four main themes. The first three themes are related to SOA, cloud-computing and blockchain architectures, and the final theme relates to novel security architectures designed explicitly for C3I systems.    

\textit{Service-oriented C3I Architecture:} C3I systems are composed of heterogeneous components (e.g., Decision support systems, Performance monitoring tools and Signal processing controllers ) that are geographically scattered and connected through diverse communication mechanisms (e.g., 4G/5G and  Wi-Fi). Therefore, Service-oriented architecture (SOA) is often considered a good fit for designing tactical C3I systems. In addition to enhanced security, SOA also helps C3I systems to achieve interoperability, scalability and platform independence among these heterogeneous system components \cite{aloisio2015tactics}. Despite these advantages, researchers have identified security limitations that impede the successful inclusion of SOA in designing C3I systems and have proposed various measures to overcome these challenges. Gkioulos and Wolthusen \cite{gkioulos2017security, gkioulos2015constraint, gkioulos2015enabling, gkioulos2016securing, gkioulos2016security} and Rigolin and Wolthusen \cite{lopes2015distributed} have proposed  security policy modeling methods (\textit{C\textsubscript{7}}) for designing secure SOA-based tactical systems. The authors have leveraged Web Ontology Language with Descriptive Logic for designing and implementing these security policies. Subsequently, Gkioulos et al. have proposed and validated a framework (\textit{C\textsubscript{8}}) for designing secure SOA-based C3I systems \cite{gkioulos2017security, gkioulos2017tactics}. They authors claimed that in addition to security, the proposed framework also provides configuration flexibility under dynamically changing network conditions, enhanced performance and improved information flow \cite{gkioulos2017security}.  

\textit{Cloud-based C3I architecture:} The adoption of cloud computing technologies offers many benefits such as convenient access, cost reduction, deployment flexibility, low maintenance and timeliness for C3I systems \cite{alghamdi2014cloud}, \cite{foster2010cloud}. However, cloud computing also introduces security concerns like data confidentiality breaches and security misconfigurations \cite{alghamdi2014cloud}. Hence, the incorporation of cloud technologies in C3I systems requires an increased focus on the secure design of C3I systems. Keeping this in focus, Jahoon et al. \cite{koo2020security} have proposed a security architecture (\textit{C\textsubscript{9}}) consisting of three layers (i.e., virtualization, physical and operating) with distinct functions to secure the overall system. The proposed architecture is built upon their findings \cite{koo2019security, koo2020security}, such as the need for server virtualization security (e.g., hypervisor and public server security) to design secure cloud-based C3I systems. Furthermore, Abdullah et al. \cite{alghamdi2014cloud} also have identified user accountability management, accurate system configuration, and uninterrupted service maintenance as some of the security necessities that must be incorporated into cloud-based C3I systems.    

\textit{Blockchain-based C3I architecture:} Blockchain technologies provide many advantages for C3I systems, including secure storage, transmission and processing of mission-critical data \cite{ozturk2019challenges}. Thus, researchers have leveraged blockchain technologies to design and implement secure C3I systems. Akter et al. have proposed a blockchain-based distributed smart-contract method for secure C3I data transmission (\textit{C\textsubscript{10}}) \cite{akter2019highly}. Through a simulation study, the authors have validated that the blockchain-based method is efficient in secure message transmission compared to real-time queuing and traditional queuing theory-based methods \cite{akter2019highly}. In a similar study, Akter et al. \cite{akter2021blockchain} have implemented a blockchain assisted cryptographic peer to peer (P2P) network for secure data communication among C3I nodes. In this approach, the researchers have used a \textit{Central Cloud Server} to store legitimate node IDs with their public keys to facilitate \textit{Local Edge Servers} in identifying intruders \cite{akter2021blockchain}. From the perspective of data sharing among trusted agencies (\textit{C\textsubscript{11}}), Razali et al. \cite{razali2021secure} have proposed a smart contract based method that supports intelligent data management (i.e., generation, edit, view and storage) and dissemination between legitimate users. This method's use of distributed and decentralized databases have provided enhanced confidentiality and availability for data from different sources (e.g., sensor, human and cyber). 

\textit{Other security architectures for C3I systems:} Here, we detail the countermeasures that present novel architectures for securing C3I systems. For example, considering architectural support for secure communication (\textit{C\textsubscript{12}}), Jin-long et al. \cite{jin2004improvement} have proposed a multi-federation architecture based on a gateway-proxy approach, facilitating inter and intra-federation secure data transmission. They have extended their approach through a double-proxy mechanism to avoid the communication overhead problem. Similarly, Alghamdi et al. \cite{alghamdi2010generic} have presented a net-centric architectures for secure C3I systems' interoperable communication mechanisms. To further enhance the security of C3I communication networks, this architecture is equipped with multiple firewalls and intruder detection systems. For sensitive data security in C3I systems (\textit{C\textsubscript{13}}), Guturu \cite{guturu2005architecture} has presented a distributed database management architecture grounded on AND-OR replication algorithm. The author has claimed that the proposed architecture improves C3I systems data storage security through high fault endurance and resilience to attacks compared to 2-phase and 3-phase commit methods. In another approach, Seungjin Baek and Young-Gab Kim have proposed a four-layer security architecture \cite{baek2021c4i} focusing on big data security in C3I systems. The four layers (i.e., application, big data platform, data, and infrastructure) provide security in data generation, data processing and data usage processes of C3I systems. We have also identified several countermeasures that focus on securing the overall C3I systems' architectures. From the perspective of securing next generation C3I systems (\textit{C\textsubscript{14}}), Perkinson \cite{perkinson2012lessons} has introduced a holistic approach for next-generation C3I systems' cyber security, consisting of four main phases, detection, correlation, visualization and response. The author has discussed key findings (e.g., the importance of automation in the decision-making process) that must be considered in future C3I systems designing based on the lessons learned from a pilot study. In a different study, Theron and Kott \cite{theron2019autonomous} have envisioned that future C3I systems should be equipped with Autonomous Cyber Defense (ACyD) methods to fight against Autonomous Intelligent Malware (AIM) attacks. They have considered Autonomous Intelligent Cyber-Defense Agents (AICAs) as a possible security mechanism that future C3I systems should be equipped with to defend themselves from AIM attacks. To develop a generic cyber-doctrine for next generation C3I systems, Ormrod and Turnbull \cite{ormrod2016cyber} have investigated the existing multinational cyber-doctrines (e.g., USA, UK, Australia and Canada) and proposed a nested domain model. The authors have stated that the proposed model provides a flexible and refined conceptual framework for future C3I cyber operations. 

In respect to supportive technologies for developing secure architectures (\textit{C\textsubscript{15}}), Alghamdi et al. \cite{alghamdi2010enhancing} have suggested that threat modeling techniques can be applied in identifying security requirements, vulnerabilities and threats associated with C3I systems. They have highlighted the necessity of embedding security at the C3I systems architectural level through employing a top-down threat modeling approach. Furthermore, Alghamdi et al. \cite{alghamdi2011proposed} have manifested a systematic approach to capture threats and formulate security defenses considering C3I systems' overall architecture. The researchers have leveraged an assurance case method and Claims Arguments Evidence tool for modeling and visualizing the security architecture, respectively. Similarly, Biagini and Corona \cite{biagini2016modelling} have leveraged the Modeling and Simulation as a Service (MSaaS) paradigm to recognize and demonstrate tools that can be incorporated into Counter Unmanned Autonomous Systems considering the security of future C3I systems.

\begin{tcolorbox}[left=0pt, top=0pt, right=0pt, bottom=0pt, boxrule=0.75pt]
\textbf{\textit{Benefits of Architectural Support for Security}}
\begin{itemize}[leftmargin=*]
    \item \small \textit{{In-depth incorporation of security into every component of C3I systems}}
    \item \textit{Instead of considering ad-hoc security mechanisms, embeds security into the architectural design of C3I systems}
    \end{itemize}
\textbf{\textit{Limitations of Architectural Support for Security}}
\begin{itemize}[leftmargin=*]
    \item \small \textit{{Time consuming to identify every security aspect to be considered in the architectural design}}
    \item \textit{Require expertise knowledge on different system components (e.g., communication networks and database modeling) to design an overall architecture considering their dependencies}
    \item \textit{Require expertise domain knowledge (e.g., cloud computing and blockchain) to incorporate security into the overall architecture }
\end{itemize}
\end{tcolorbox}

\subsubsection{Attack Impact Analysis}

Attack impact analysis refers to the study of adverse impacts of attack vectors executed on C3I systems. The attack impact analysis helps in predicting possible consequences of attack vectors, which also highlights security loopholes in C3I systems design. The following countermeasures, \textit{C\textsubscript{16} to C\textsubscript{21}}, present different approaches for attack impact analysis in C3I systems.

Fang et al. \cite{fang2015research} have evaluated two types of cyber impacts (i.e., system structure damaged effect and structure efficiency reduction effect) for C3I systems. The authors have presented mathematical modeling and simulation study (\textit{C\textsubscript{16}}) for evaluating attack impacts through different statistical parameters such as \textit {connectivity rate} and \textit {information efficient}. Similarly, Jinfeng et al. \cite{jinfeng2017effects} (\textit{C\textsubscript{17}}) have used information flow parameters, such as intelligence, command-and-control, and collaboration information flow parameter, to calculate the C3I system structure effects (i.e., damaged effect and fallback effect). Focusing on metrics-driven cyber impact analysis, Bernier et al. \cite{bernier2012metrics} have presented a metrics framework (\textit{C\textsubscript{18}}) to assess the impact of cyber operations in C3I systems. The researchers have mainly considered three metrics (i.e., measures of force effectiveness, command-and-control effectiveness and performance) along with the US DoD framework and goal-question-metric paradigm \cite{basili1992software} to evaluate cyber impacts in C3I systems.

Since simulation studies provide cost, time and power efficient ways for analyzing attack impacts on C3I systems, researchers have conducted simulation studies to investigate cyber impacts on C3I systems. For example, Mursia et al. \cite{mursia2011simulation} (\textit{C\textsubscript{19}}) have executed three attack vectors (i.e., Eavesdropping (\textit{A\textsubscript{11}}), Flooding Attack (\textit{A\textsubscript{12}}), and Jamming Attack (\textit{A\textsubscript{13}})) on a C3I MANET environment created in EXata/Cyber emulator. The researchers have analyzed cyber impacts in terms of network parameters: throughput, latency and jitter. Similarly, Morton et al. \cite{morton2012simulation} have conducted a simulation study (\textit{C\textsubscript{20}}) for executing military cyber operations. The authors have reported several adverse effects of cyber-attacks such as unauthorized use, interception, and degradation of C3I systems. Another simulation approach, called cyber simulation terrain (\textit{C\textsubscript{21}}) \cite{o2015cyber}, is based on ontological network representation for modelling cyber assets and systems of computer networks (e.g., C3I networks). The authors have investigated the impacts of heartbleed vulnerability (\textit{V\textsubscript{11}}) in simulated computer network operations.

\begin{tcolorbox}[left=0pt, top=0pt, right=0pt, bottom=0pt, boxrule=0.75pt]
\textbf{\textit{Benefits of Attack Impact Analysis}}
\begin{itemize}[leftmargin=*]
    \item \small \textit{{Helps executives and stakeholders to assess and visulaize cyberscurity risk in C3I systems} }
    \item \textit{Facilitates C3I systems' operators to estimate damages in case of successful cyber-attacks}
    \item \textit{Provides multiple viewpoints and parameters to investigate adverse effects of cyber-attacks}
\end{itemize}
\textbf{\textit{Limitations of Attack Impact Analysis}}
\begin{itemize}[leftmargin=*]
    \item \small \textit{{Lacking in-depth knowledge of each C3I component and its operating environment to assess cybersecurity risk of C3I systems}}
    \item \textit{Inability to replicate cyber-attacks as its original incident, which may result in inaccurate impact analysis and quantification}
\end{itemize}
\end{tcolorbox}

\subsection{Operational Countermeasures}

Operational countermeasures refer to the security mechanisms proposed for protecting C3I systems during tactical operations. The existing literature has reported intrusion detection mechanisms, cryptographic techniques and access control approaches for securing C3I operations. In the following, we describe the proposed countermeasures along with their benefits and limitations.

\subsubsection {Intrusion Detection}

Intrusion detection mechanisms aim at detecting cyber-attacks on C3I systems. When it comes to cybersecurity, time is a critical factor. The sooner an attack can be detected, the sooner the attack will be contained/mitigated. According to \cite{GDPR2017}, timely detection of an attack can reduce the success rate of an attack by 97\%. However, noise and limited resources of communication networks damage the effectiveness of intrusion detection mechanisms \cite{arshad2020review}, which leads to false-alarm generation and communication overhead during time-critical C3I operations. Tactical domains, such as disaster management and rescue operations, cannot endure such anomalies in C3I operations due to their sensitive nature. Therefore, researchers have proposed the following effective intrusion detection approaches, \textit{ C\textsubscript{22} to C\textsubscript{26}}, to ensure the prompt detection of intrusions in C3I tactical systems.

Shaneman et al. \cite{shaneman2007enhancing} have proposed a cost-effective mechanism, called intrinsic fiber monitoring (\textit{C\textsubscript{22}}), to detect intrusions in fiber-optic cables used for data transmission in C3I networks. The researchers have experimentally proved that the intrinsic fiber monitors neither disturb C3I network resources (e.g., bandwidth) nor generate false alarms while detecting malicious activities (e.g., tampering attack (A\textsubscript{6})) in C3I systems. For early detection of intrusions, Manes et al. \cite{manes2003identifying} have presented an attack management framework (\textit{C\textsubscript{23}}) that combines network vulnerability models (i.e., vulnerability analysis) with attack models (i.e., attack trees) to detect the early signs of multi-stage attack vectors in C3I networks. The proposed architecture also facilitates the visualization of intruders activities in C3I networks, which helps security experts in preparing action plans against cyber adversaries in real-time.

To discover malicious activities of intruders in tactical MANETs, a light-weight C3I monitoring system (\textit{C\textsubscript{24}}) based on Multi-Instance Multi-Label learning protocols has been proposed in \cite{roopa2018intelligent}, which causes less computational and communication overhead in C3I networks as compared to the existing solutions such as watchdog scheme and digital signatures. Concerning the security of SOA based C3I systems, Jorma and Jan \cite{jormakka2009intruder} have developed an intruder detection system architecture (\textit{C\textsubscript{25}}) underpinned by four design patterns (i.e., linked list, sandbox, event bus, and monitor). The proposed architecture uses active and passive monitoring techniques, such as detection and alarm rules, for detecting intruders in SOA based C3I systems. Another mechanism, presented by Kwon et al. \cite{kwon2019identification}, proposes an AI-based C3I device identifier (\textit{C\textsubscript{26}}) based on Wiener filters, MIRtoolbox and ML models (i.e., Support vector machine). The presented mechanism is used in digital forensics for detecting malicious activities (e.g., data deletion and modification) in the stored data of C3I databases.

\begin{tcolorbox}[left=0pt, top=0pt, right=0pt, bottom=0pt, boxrule=0.75pt]
\textbf{\textit{Benefits of Intrusion Detection Mechanisms}}
\begin{itemize}[leftmargin=*]
    \item \small \textit{{Cost effective and resource efficient mechanisms for detecting cyber-attacks in resource constrained tactical domains}}
    \item \textit{Elimination of adversarial attacks from its initial penetration point}
    \item \textit{Possibility of predicting and taking proactive measures against potential future attacks}
\end{itemize}
\textbf{\textit{Limitations of Intrusion Detection Mechanisms}}
\begin{itemize}[leftmargin=*]
    \item \small \textit{{Due to inaccurate inferences (false positives), even legitimate users might be obstructed from accessing to services}}
    \item \textit{Need of human intervene for correcting false-classifications and incorporating new or existing threat variations }
\end{itemize}
\end{tcolorbox}

\subsubsection {Cryptography}

Cryptography is a process of encoding and decoding sensitive information. Considering the criticality of C3I tactical operations (Section \ref{introduction}), cryptography has been used in C3I systems for preventing unauthorized disclosure of sensitive information \cite{manuel2021implementing}. For this purpose, cryptographic functions, along with their respective cryptographic keys, are used to encrypt and decrypt tactical information stored and transmitted among C3I systems. The cryptographic mechanisms, \textit{ C\textsubscript{27} to C\textsubscript{34}}, are mostly focused on cryptographic key management in tactical environments for securing real-time data transmission and storage in C3I systems.

Cryptographic key management refers to the process of generating, exchanging, storing, deleting and replacing cryptographic keys in tactical C3I systems. In general, cryptographic key management solutions are costly in terms of both computational and network resources usage \cite{matt2004toward}, which leads to communication overhead and delay in C3I tactical operations. Therefore, the existing literature presents efficient ways for managing cryptographic keys in C3I systems. For example, Kang et al. \cite{kang2006towards} (\textit{C\textsubscript{27}}) have used symmetric-key algorithms to generate cryptographic keys of different lengths according to the severity levels of security risks in C3I systems. The proposed mechanism provides both security and optimal QoS of C3I systems simultaneously in tactical domains. Focusing on energy-efficiency, Chudzikiewicz et al. \cite{chudzikiewicz2019procedure} have introduced a secure key management mechanism (\textit{C\textsubscript{28}}) in which a central node uses a Quantum Random Number Generator to generate cryptographic keys, and distribute them in C3I networks through secure session management among C3I nodes. Another key management process (\textit{C\textsubscript{29}}), proposed in \cite{alghamdi2011satellite}, uses a HenLog- Random Key Generator for secure key generation, which provides a high level of confusion and diffusion in the encryption process of satellite images used in C3I systems. Furthermore, Furtak et al. \cite{furtak2016security} have presented a time and memory efficient approach for key management (\textit{C\textsubscript{30}}) based on symmetric-key cryptography to secure a network access layer of C3I networks. The authors have also proposed an asymmetric cryptographic approach \cite{furtak2019security} (\textit{C\textsubscript{31}}), which is more time-efficient, for key management and secure data transmission in C3I systems. To secure the cryptographic key distribution process, Matt et al. \cite{matt2004toward} have presented a non-interactive identity-based key-sharing scheme (\textit{C\textsubscript{32}}) among C3I nodes. The proposed mechanism consumes less computational power and network resources (e.g., bandwidth) to generate and share cryptographic keys in C3I systems.

Apart from the abovementioned key management solutions, the state-of-the-art also reports data encryption techniques for sensitive data protection in tactical environments. For instance, Romero-Mariona et al. \cite{romero2013cealician} have introduced an encryption device called \textit{CEALICIAN} (\textit{C\textsubscript{33}}) that uses NSA-Suite B cryptographic algorithms to encrypt or decrypt sensitive information at a high data rate in real-time. The \textit{CEALICIAN} focuses on less power consumption and form factor to address the size, weight and power requirements of C3I systems. Furthermore, Abdullah et al. \cite{alghamdi2010evaluating} (\textit{C\textsubscript{34}}) have qualitatively evaluated three data encryption methods (i.e., advanced encryption standards, Ron Rivest, Adi Shamir, and Leonard Adleman, and quantum encryption) for C3I systems. Based on their research findings and the criticality of C3I systems, it has been concluded that the advanced encryption standards are more suitable for C3I systems compared to other examined encryption methods.

\begin{tcolorbox}[left=0pt, top=0pt, right=0pt, bottom=0pt, boxrule=0.75pt]
\textbf{\textit{Benefits of Cryptographic Mechanisms}}
\begin{itemize}[leftmargin=*]
    \item \small \textit  {{Provides computational and network resource efficient mechanisms for resource constrained tactical environments}}
    \item \textit{Possibility of adopting into different types of C3I components such as MANET and Web services}
   % \item \textit{Time-efficient cryptographic approaches facilitate achieving tactical goals in time-critical C3I missions}
    \end{itemize}

\textbf{\textit{Limitations of Cryptographic Mechanisms}}
\begin{itemize}[leftmargin=*]
    \item \small \textit{{Need of specialized hardware or complex algorithms}}
    %\item \textit{Expensive computational processing and memory}
    \item \textit{Added overhead on resource scarce devices such as sensors and wireless transceivers}
\end{itemize}
\end{tcolorbox}

\subsubsection{Access Control}

Access control refers to security mechanisms that allow authorized users, such as field commanders and decision-makers, to access C3I systems. As described in Section \ref{overview}, C3I systems follow a strict chain-of-commands that has multiple users with different access rights to execute C3I operations. Therefore, access control mechanisms are necessary for C3I systems. Since tactical missions success, lives and property largely depend upon access control mechanisms \cite{maschino2003access}, C3I systems should employ reliable, flexible and efficient access control approaches for preventing unauthorized access in tactical domains. Therefore, researchers focus on developing effective access control mechanisms, \textit{ C\textsubscript{35} to C\textsubscript{40}}, for tactical C3I systems.

Zachary \cite{zachary2003decentralized} has presented a decentralized approach (\textit{C\textsubscript{35}}) based on one-way accumulators for authenticating legitimate nodes in distributed C3I networks. The proposed mechanism securely manages a node membership list for all C3I nodes, while utilizing less computational and network resources (e.g., power and bandwidth), in dynamic C3I networks. Similarly, a cost-effective and power-efficient technology, called digital-ultrasonic sensor network technology \cite{forcier2003wireless} (\textit{C\textsubscript{36}}), has been proposed for preventing unauthorized access in C3I systems. The digital ultrasonics outperforms the existing alternatives such as infrared, Bluetooth, and ultra-Wide Band etc. To ensure access control in horizontal and vertical chain-of-commands in C3I systems, Jarmakiewicz et al. \cite{jarmakiewicz2015design} have proposed multi-level security and multiple independent level security mechanisms (\textit{C\textsubscript{37}}) based on XACML and WEB services. The proposed mechanisms are time efficient in authorizing legitimate users for operating C3I systems. Similarly, Maule \cite{maule2005enterprise} has presented a multi-level security framework, known as multi-layered ontology security model (\textit{C\textsubscript{38}}), to safeguard military information system data by using different authorization levels. Another multi-layer model (\textit{C\textsubscript{39}}), reported in \cite{rezakhani2018novel}, is used to implement authentication, authorization and accountability in C3I systems. The multi-layer approach provides a road map for managing access in integrated services (e.g., SOA architecture) based on operational requirements. For efficient access control in a C3I command hierarchy, Maschino \cite{maschino2003access} has compared different combinations (i.e., centralized and decentralized) of Role-Based Access Control (RBAC) mechanisms (\textit{C\textsubscript{40}}) for centralized and decentralized C3I systems. Consequently, the author has recommended a hybrid RBAC mechanism for ensuring role-based authentication in C3I systems.

\begin{tcolorbox} [left=0pt, top=0pt, right=0pt, bottom=0pt, boxrule=0.75pt]
\textbf{\textit{Benefits of Access Control Mechanisms}}
\begin{itemize}[leftmargin=*]
    %\item \small \textit{{Reduces the unauthorized access to C3I system functionalities}}
    \item \small \textit{Restrict the unauthorized access, origination, and retention of C3I sensitive data}
    \item \textit{Mechanisms such as multi-level security and RBAC, provide strict access control at the each level of C3I chain-of-command}
\end{itemize}
\textbf{\textit{Limitations of Access Control Mechanisms}}
\begin{itemize}[leftmargin=*]
    \item \small \textit{{Implementation difficulties in widely distributed C3I systems across tactical domains}}
    \item \textit{Misconfigured access control measures provide false sense of security}
    \item \textit{Increased complexity of user management when the system expands}
\end{itemize}
\end{tcolorbox}

\section{Discussion} \label{discussion}

In this section, we analyze our research findings reported in Sections \ref{vulnerabilites}, \ref{attacks}, and \ref{countermeasures}. Based on our analysis, we develop a distinctive relationship between attack vectors, vulnerabilities and countermeasures. Furthermore, we propose future research directions to guide researchers and practitioners in advancing the literature on cybersecurity of C3I systems. 

\subsection{Mapping C3I Attack Vectors to Vulnerabilities}

\begin{figure}[!bp]
  \centering
  \includegraphics[trim=0 280 0 0,clip, scale = 0.4]{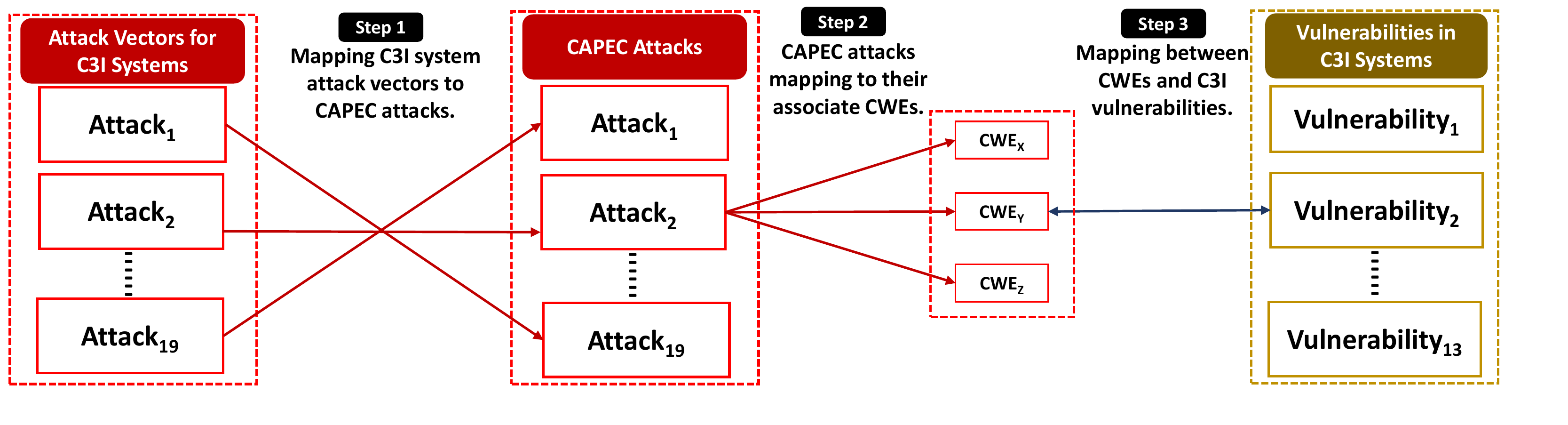}
  \caption{The methodology followed for mapping C3I attack Vectors and vulnerabilities. CAPEC (Common Attack Pattern Enumeration and Classification) and CWE (Common Weakness Enumeration) }
  \label{fig:mapping-methodology} 
\end{figure}

 While Sections \ref{vulnerabilites} and \ref{attacks} respectively describe the individual security vulnerabilities and attack vectors, they do not show any association or relationship (i.e., how a particular attack vector can exploit C3I system vulnerabilities). Therefore, to support practitioners and developers in building secure C3I systems, we articulated a mapping between C3I systems' security vulnerabilities and attack vectors. Our mapping is based on the data extracted from Common Attack Pattern Enumeration and Classification (CAPEC)\footnote{https://capec.mitre.org/} and Common Weakness Enumeration (CWE)\footnote{https://cwe.mitre.org/} databases. CAPEC provides a dictionary of attack vectors used to exploit known vulnerabilities. On the other hand, CWE specifies the list of known hardware and software vulnerabilities.

 As Figure~\ref{fig:mapping-methodology} illustrates, we initially mapped the identified C3I attack vectors with the CAPEC attacks and extracted their associated CWE numbers. Then, we mapped corresponding CWEs to the C3I security vulnerabilities that were extracted from the surveyed studies. For example, firstly, we mapped \textit{Brute Force Attack (A\textsubscript{1})} to \textit{CAPEC-112: Brute Force Attack} based on the information provided in the surveyed studies \cite{kang2006towards} and CAPEC database \cite{BruteForceCAPEC}, respectively. Secondly, \textit{CWE-521: Weak Password Requirements} \cite{WeakPasswordCWE} was identified as a corresponding CWE for the above mentioned CAPEC attack. Thirdly, we linked this CWE with \textit{Insecure Configuration Stores (V\textsubscript{7})} since the use of weak passwords \cite{alghamdi2010enhancing} is one of the root causes resulting into insecure configuration stores. Since all our identified attack vectors and vulnerabilities are not available in CAPEC and CWE, we consulted external sources (e.g., white papers and blogs) to make this mapping further comprehensive by identifying other possible associations between identified vulnerabilities and attack vectors. For example, we mapped \textit{Flooding Attack (A\textsubscript{12})} with \textit{Insufficient Logging and Monitoring (V\textsubscript{1})} vulnerability since the lack of network traffic monitoring allows attackers to flood C3I nodes with a large volume of malicious data and obstruct the communication between legitimate C3I nodes \cite{manna2012review}.

 \begin{figure}[!tbp]
  \centering
  \includegraphics[trim=130 485 20 70,clip, scale = 0.95]{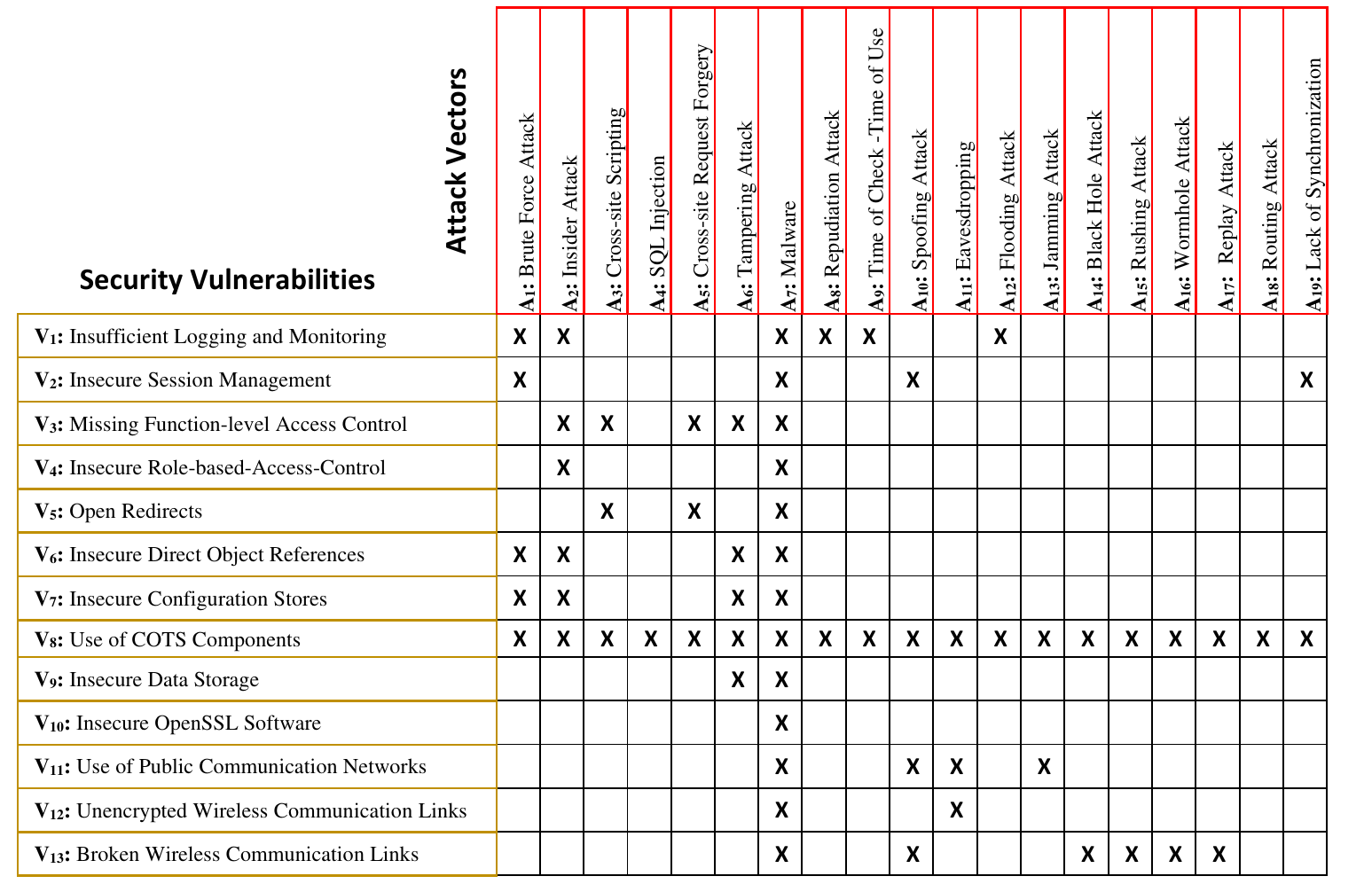}
  \caption{Mapping between identified C3I attack vectors and vulnerabilities}
  \label{fig:attack-vulnerabilities} 
\end{figure}

As Figure~\ref{fig:attack-vulnerabilities} presents, a vulnerability can be exploited by multiple attacks (e.g., vulnerability V\textsubscript{1} can be exploited by attack vectors A\textsubscript{1}, A\textsubscript{2}, A\textsubscript{8} and A\textsubscript{12}) and an attack vector can exploit multiple vulnerabilities (e.g., attack vector A\textsubscript{1} can exploit vulnerabilities V\textsubscript{1}, V\textsubscript{2}, V\textsubscript{6} and V\textsubscript{7}). Thus, it reflects the necessity of employing investigative mechanisms such as threat modeling to identify all possible associations between vulnerabilities and attack-vectors before designing security defenses. For example, to prevent \textit{Brute Force Attack (A\textsubscript{1})} execution in C3I systems, developers must not only consider implementing cryptographically strong user credential storage, but also incorporating adequate intrusion detection and prevention mechanisms. As shown in Figure \ref{fig:attack-vulnerabilities}, the use of COTS component is exploitable by all identified attack vectors. This is because the use of vulnerable COTS components in C3I systems expose the system to several attack vectors since these third-party components can contain intentional (e.g., backdoors) or unintentional (e.g., inadequate testing) security flaws. Thus, it emphasizes conducting rigorous security and quality testing before integrating these components into C3I systems. Moreover, we recommend that such testing and quality checks must be incorporated as a mandatory procedure in the designing phase of the C3I system development process. There are numerous ways malware can exploit vulnerabilities, both directly and indirectly, through different penetration methods (e.g., over the network and downloading from the internet \cite{ye2017survey}). Therefore, in this mapping, we consider that malware can exploit all the reported C3I vulnerabilities. For example, malware can capitalize on \textit{V\textsubscript{7}: Insecure data storage} and access C3I systems' security-critical data. On the other hand, malware can exploit \textit{V\textsubscript{1}: Insufficient logging and monitoring} vulnerability and spy on C3I operations. Therefore, we emphasize the requirement of employing malware detection methods (e.g., AI/ML and pattern mining \cite{ye2017survey}) in C3I systems.

% We acknowledge that these attack vectors can also exploit other vulnerabilities since a single attack vector can exploit multiple vulnerabilities and vice versa. Therefore, we would like to highlight the need for future investigations on comprehensive and systematic mappings between C3I systems' vulnerabilities and attack vectors. 

\subsection{Mapping C3I Attack Vectors to Countermeasures}

\begin{figure*}[!htbp]
  \centering
  \includegraphics[trim=55 124 0 71,clip, scale = 0.82]{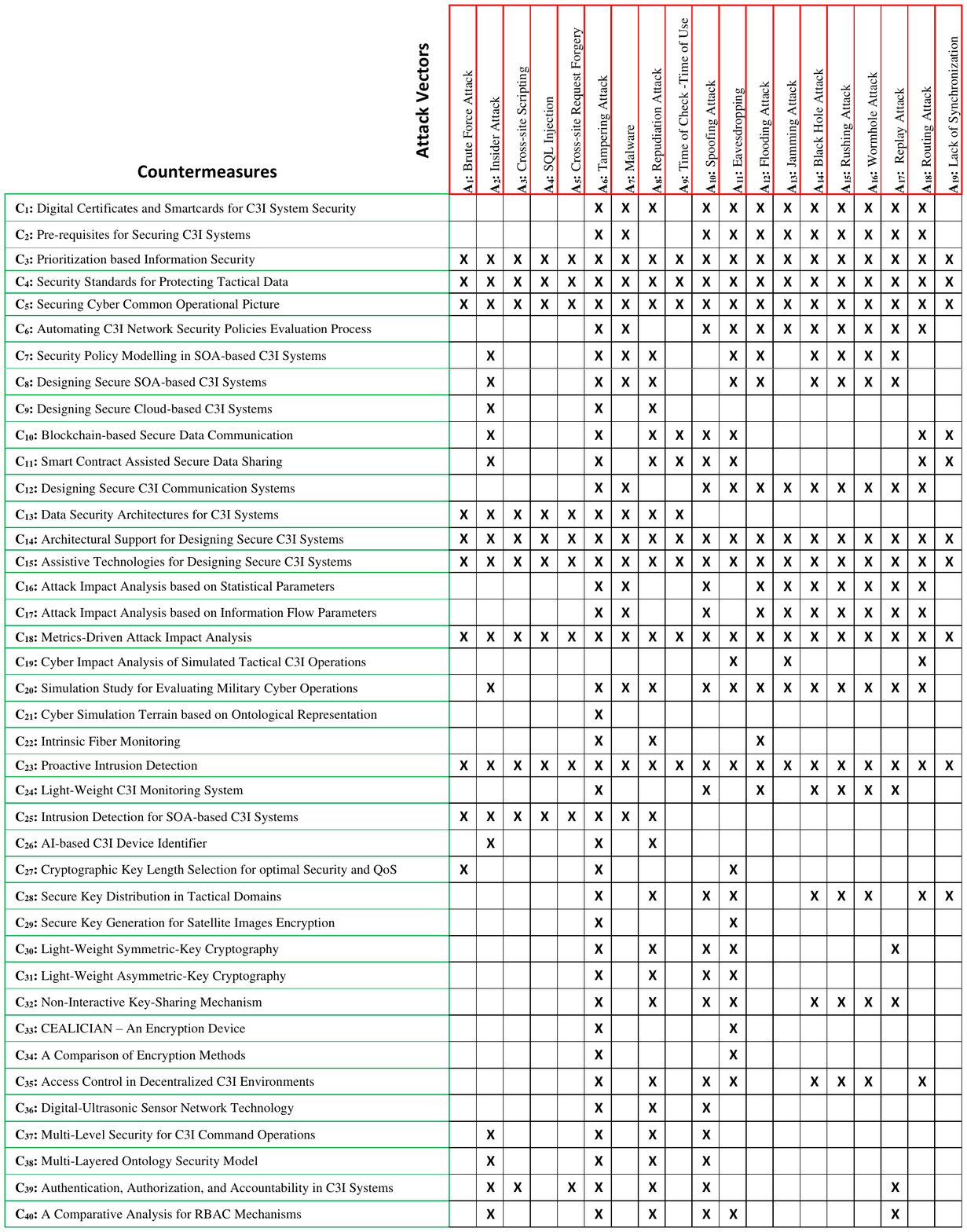}
  \caption{Mapping between C3I attack vectors and countermeasures.
  \textit{C\textsubscript{1}} to \textit{C\textsubscript{21}} applicable for development phase, \textit{C\textsubscript{22}} to \textit{C\textsubscript{40}} applicable for operational phase of C3I systems}
  \label{fig:attack-countermeasures} 
\end{figure*}

As discussed in the previous sub-section, the existing literature does not report a clear relationship between attack vectors, vulnerabilities, and countermeasures. Therefore, similar to attacks-vulnerabilities mapping, we map attack vectors to countermeasures for demonstrating which countermeasures can defend C3I systems against which attack vectors. Such a relationship accumulates possible countermeasures for an attack vector, which facilitates researchers and practitioners in identifying solutions against an attack vector. Since attack vectors and countermeasures can be linked in several ways, we focus on developing an evident relationship between them. For this purpose, we have developed a two-step approach to gather reported and unreported distinct attacks-countermeasures relationships. In the first step, we have identified the relationship between attack vectors and countermeasures directly from the existing literature. For example, the \textit{variable cryptographic key length (C\textsubscript{27}}) provides protection against \textit{brute force attack (A\textsubscript{1}}) as reported in \cite{kang2006towards}. Similarly, the \textit{AI-based C3I device identifier (C\textsubscript{26}}) detects sensitive data manipulation (\textit{A\textsubscript{6}}) in C3I databases \cite{kwon2019identification}. In the second step, we have thoroughly analyzed the attack vectors and countermeasures to identify relationships between them. For instance, the \textit{CEALICIAN (C\textsubscript{33}}) encrypts and decrypts sensitive information transmitted through C3I communication links, which hinders the \textit{eavesdropping attack (A\textsubscript{11}}). Similarly, the \textit{non-interactive key sharing scheme (C\textsubscript{32}}) mitigates the chances of \textit{tampering attack ({A\textsubscript{6}})} during data transmission. Therefore, we have mapped the \textit{C\textsubscript{33}} and \textit{C\textsubscript{32}} to the \textit{A\textsubscript{11}} and \textit{A\textsubscript{6}} respectively. In this way, we have created a distinct relationship between attack vectors and countermeasures through our two-step approach. Figure~\ref{fig:attack-countermeasures} presents the mapping of the attack vectors to their corresponding countermeasures.

We observe that the countermeasures concerning the development phase of C3I systems (i.e., \textit{C\textsubscript{1}} to \textit{C\textsubscript{21}}) provide holistic recommendations for securing C3I systems from most, if not all, attack vectors. For example, the \textit{designing secure C3I communication systems (C\textsubscript{12})}, if implemented rigorously, prevents several attack vectors such as  \textit{A\textsubscript{6}}, \textit{A\textsubscript{7}}, \textit{A\textsubscript{10}}, \textit{A\textsubscript{11}}, \textit{A\textsubscript{12}}, \textit{A\textsubscript{13}}, \textit{A\textsubscript{14}}, \textit{A\textsubscript{15}}, \textit{A\textsubscript{16}}, \textit{A\textsubscript{17}}, and \textit{A\textsubscript{18}}. Similarly, blockchain based design patterns (i.e., \textit{C\textsubscript{10}} and \textit{C\textsubscript{11}}) not only prevent attack vectors targeting C3I communication links (i.e., \textit{A\textsubscript{10}} and \textit{A\textsubscript{11}}), but also protect against \textit{repudiation attack (A\textsubscript{8})} in C3I systems. Therefore, the countermeasures (i.e., \textit{C\textsubscript{1}} to \textit{C\textsubscript{21}}) related to the development category provide solution against most of the attack vectors as depicted in Figure~\ref{fig:attack-countermeasures}. However, organizations employing C3I systems cannot implement security measures extensively at once due to their limited expertise, financial and technological resources. Therefore, we emphasize the need of prioritizing the security measures while developing C3I systems. For this purpose, we assert that researchers and practitioners should first analyze C3I systems' operational and environmental conditions and then implement ad hoc countermeasures accordingly during the development of C3I systems.

Regarding the operational countermeasures (\textit{C\textsubscript{22}} to \textit{C\textsubscript{40}}), we observe a lack of interoperability between security tools providing intrusion detection, cryptography and access control capabilities in C3I systems. For example, the detective countermeasure \textit{AI-based C3I device identifier (C\textsubscript{26})} only detects the \textit{tampering attack A\textsubscript{6}} execution, but, to prevent \textit{A\textsubscript{6}}, the preventive countermeasure \textit{non-interactive key sharing mechanism (C\textsubscript{32}}) needs to be implemented. On the contrary, the preventive countermeasures (e.g., C\textsubscript{27} to C\textsubscript{40}) take more time in the absence of detective countermeasures (e.g., \textit{C\textsubscript{22}} to \textit{C\textsubscript{26}}) for hampering the execution of attack vectors \cite{GDPR2017}. Therefore, we assert that researchers should formulate novel defense mechanisms that should not only be able to integrate security tools but also be flexible enough to accommodate new security tools for efficiently detecting and preventing attack vectors in time-critical C3I systems.

\subsection{Future Research Areas}
Based on the findings presented in Sections \ref{vulnerabilites}, \ref{attacks}, and \ref{countermeasures}, we propose the following future research directions for advancing the state-of-the-art on the cybersecurity of C3I systems.
%Several research works are being done for improving the security of C3I systems. Some researchers are making improvements in the existing countermeasures (e.g., \cite{kwon2019identification}) while others are exploring new, innovative solutions to secure C3I operations (e.g., \cite{theron2019autonomous}, \cite{ormrod2014coordination}). We have identified following emerging research areas in the development of secure C3I systems. 

\subsubsection{Secure SOA adaptation for C3I systems}
As discussed in Section \ref{overview}, a C3I system is an integration of multiple heterogeneous systems. Therefore, traditional monolithic architecture is not a viable choice for designing large scale C3I systems \cite{namiot2014micro}. Driven by the necessity to adopt fitting architecture patterns, researchers have considered SOA to design C3I systems (e.g., Deployable Joint C3I System \cite{malik2012application}). However, SOA-based systems suffer from various security vulnerabilities and attack vectors \cite{lowis2009classification, masood2013cyber}, raising many security challenges when adopting into security-critical tactical systems such as C3I systems. For example, SOA lacks continuous monitoring, and intrusion detection mechanisms by design \cite{jormakka2009intruder}. Hence, SOA-based C3I systems are vulnerable to numerous attack vectors (e.g., \textit{Brute Force Attack (A\textsubscript{1})}, \textit{Insider Attack (A\textsubscript{2})} and \textit{repudiation attack (A\textsubscript{8})}). Therefore, considering the contemporary efforts for addressing the security challenges in adapting SOA to C3I systems (Section \ref{security_by_design}), it is evident that the secure adaptation of SOA into C3I systems is an upcoming research field. Therefore, we assert that the cybersecurity of SOA-based C3I systems must be critically considered, and novel security measures should be embedded into the design and implementation of these systems. Furthermore, it is also observed that research on the SOA-based C3I domain require comprehensive evaluation efforts, including experiments on real-world implementations, as some of the existing studies are limited to conceptual solutions only.

%Aloisio et al. have presented a project, namely, TACTICS, which discusses the adaptation of SOA into the tactical domain with the focus of integrating military-grade security into tactical C3I systems [S76]. Based on this project, researchers have explored various aspects related to security policy modelling in SOA tactical systems \cite{gkioulos2017security, gkioulos2015constraint, gkioulos2015enabling, gkioulos2016securing, gkioulos2016security, lopes2015distributed, gkioulos2017tactics}. Furthermore, in a separate study, Jormakka and Lucenius \cite{jormakka2009intruder} and Jarmakiewicz and Podlasek \cite{jarmakiewicz2015design} have investigated incorporating intrusion detection techniques and multilevel security models into SOA-based C3I systems. 

%Considering the increased interest of researchers and practitioners, it is evident that the secure adaptation of SOA into tactical C3I systems is an upcoming research field. Therefore, we assert that the cybersecurity of SOA-based C3I systems must be critically considered, and novel security measures should be embedded into the design and implementation of these systems. Furthermore, it is also observed that research on the SOA-based C3I domain require comprehensive evaluation efforts, including experiments on real-world implementations, since some of the existing studies are limited to conceptual solutions only. 

\subsubsection{Lightweight security measures} C3I systems are usually operated in resource-constrained environments. Therefore, C3I systems are designed according to certain bandwidth, energy, and memory specifications. This implies that each component of a C3I system needs to be operated with limited resources. This holds true for the security measures (e.g. intrusion detection, cryptographic mechanisms, and access control) incorporated in C3I systems. If security measures consume more computational power, on-board storage or network resources for securing C3I systems, this results in affecting strategic operations of C3I systems. Though researchers have proposed resource efficient solutions that separately consider power \cite{forcier2003wireless}, memory \cite{furtak2016security}, computing resource \cite{kang2006towards}, and network resource \cite{shaneman2007enhancing} efficiency, we observe a lack of holistic and optimal security measures similar to the \textit{CEALICIAN} (\textit{C\textsubscript{33}}) that not only provides power efficiency, but also considers size and weight requirements of C3I systems while encrypting and decryption sensitive information. Therefore, we assert that researchers should focus on developing light-weight security measures that consider most, if not all, parameters of lightweight cybersecurity.

%that are often termed as SWaP requirements. This implies that each component of a C3I system needs to operate in an environment constrained by size, weight, and power. This holds true for the security measures (e.g. IDS, firewall, and malware detector) incorporated in a C3I system. This results in affecting other operations of the system. In other words, these security measures are not suppose to eat up all resources of the C3I system. That is why, 13 papers in our review have highlighted this issue. We endorse the highlighted point that the security measures incorporated should be computationally light-weight so that these security measures do not end up significantly affect other operations of a C3I system.

\subsubsection{Secure cloud-based C3I architecture}
Although cloud computing generally is a widely used technology, its incorporation into the C3I domain is quite new. As discussed in Section \ref{security_by_design}, cloud computing provides many benefits (e.g., cost reduction, convenient access, and deployment flexibility) for improving the efficiency and operability of C3I systems in tactical domains \cite{alghamdi2014cloud}. Considering these advantages, the United States has developed cloud-based C3I systems for its national defense and public sector, and Korea is shifting its national defense infrastructure to cloud-based systems \cite{koo2019security}. However, the integration of cloud computing with C3I systems possesses many security challenges (e.g., data confidentiality, integrity and server visualization), as reported in \cite{alghamdi2014cloud, koo2019security, koo2020security}. For example, the confidentiality of data can be breached in the case of the public cloud, where multiple users share the same infrastructure. Similarly, cloud service models such as PaaS and IaaS allow users to install their own software; this privilege can damage the integrity of sensitive data \cite{alghamdi2014cloud}. We identified only one attempt (i.e., \cite{koo2020security}) to design secure cloud-based C3I systems from the surveyed articles. Since the cybersecurity and efficiency of C3I systems cannot be considered orthogonal, therefore, we assert that researchers need to focus on addressing the security challenges in developing cloud-based C3I systems to enhance the efficiency and cybersecurity of C3I systems simultaneously.

\subsubsection{Blockchain for securing C3I infrastructure}
Blockchain technology provides strict authentication, secure distribution of storage facilities and computational resources across a C3I tactical domain \cite{akter2019highly, ozturk2019challenges}, which avoids the chances for a single point of failure in C3I systems. However, we have found significantly low research interest in incorporating blockchain technologies into C3I systems, compared to other domains, such as IoT \cite{cho2019survey} and economic systems \cite{zargham2018state}. The approaches proposed in reviewed studies are limited to secure data transmission \cite{akter2019highly, akter2021blockchain} and data sharing \cite{razali2021secure} in tactical C3I systems. Therefore, it is evident that blockchain technology is relatively new in C3I domains. Considering the importance of the blockchain in tactical domains \cite{kulshrestha2018military, mcabee2019military}, we assert that researchers should explore new directions for incorporating blockchain technology for the cybersecurity of C3I systems.

% \begin{figure}[!hbp]
% \centering
% \begin{subfigure}{.45\textwidth}
%   \centering
%   \includegraphics[trim=0 293 510 0,clip, scale = 0.45]{samples/images/Evaluation.pdf}
%   \caption{Evaluation methods}
%   \label{fig:evaluation} 
% \end{subfigure}%
% \begin{subfigure}{.45\textwidth}
%   \centering
%   \includegraphics[trim=5 380 498 0,clip, scale = 0.45]{samples/images/Tools.pdf}
%   \caption{Extracted software and hardware tools.}
%   \label{fig:tools}
% \end{subfigure}
% \caption{Workload run-times with Hadood, Spark and Flink frameworks}
% \end{figure}

\subsubsection{AI/ML for the cybersecurity of C3I systems}
The use of AI/ML for cybersecurity is increasingly explored. For instance, in contrast to signature-based intrusion detection, ML-based intrusion detection is widely investigated \cite{kilincer2021machine}. Similarly, AI/ML approaches are used for detecting data exfiltration and automating response process to cyber incidents \cite{ullah2018data, sabir2021machine, islam2020architecture}. However, we find little evidence (only 3 papers) in support of using AI/ML for enhancing the cybersecurity of C3I systems. For example, Kwon et al. \cite{kwon2019identification} leveraged supervised learning for classifying C3I mobile devices into trusted and untrusted categories. Similarly, among the surveyed paper, we found that the authors in \cite{roopa2018intelligent} used multi-instance-multi-label learning models for detecting intrusion attempts on C3I systems. Furthermore, Kott et al. \cite{kott2018autonomous} have presented an initial reference architecture based on the intelligent-agent technology (i.e., Autonomous Intelligent Cyber-defense Agent) for securing C3I systems. As we found only three papers leveraging AI/ML for securing C3I systems, we believe that there is room for further exploration along the side of using AI/ML for C3I security. In this regard, we assert that researchers should explore the use of AI/ML for deep packet inspection, anomaly detection, APT detection, and detecting data exfiltration attempts on C3I systems.

\begin{figure}[!tp]
\begin{minipage}{.43\textwidth}
\centering
  \centering
  \includegraphics[trim=0 285 510 0,clip, scale = 0.4]{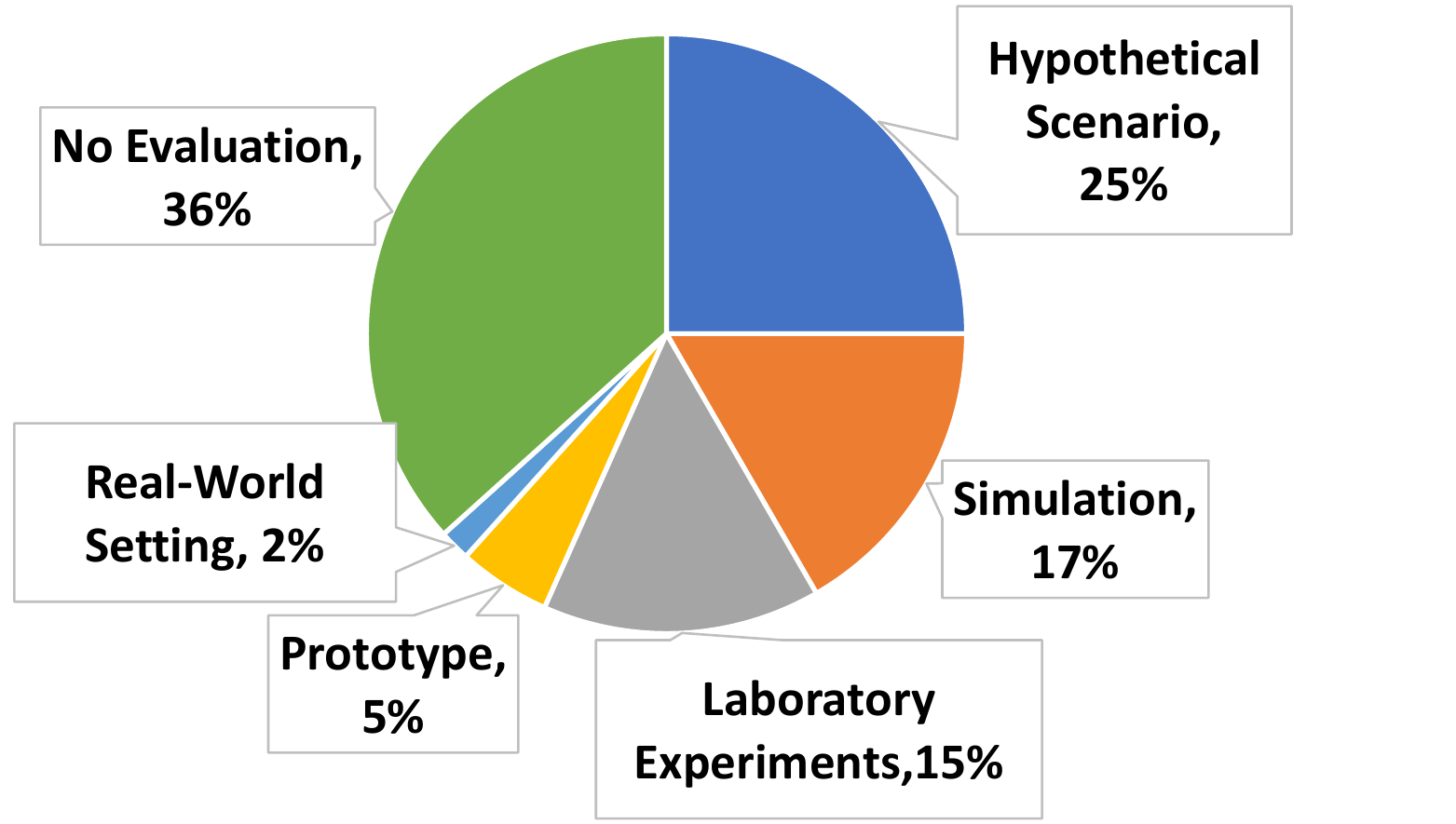}
  \caption{Evaluation methods}
  \label{fig:evaluation} 
\end{minipage}
\begin{minipage}{.47\textwidth}
 \centering
  \includegraphics[trim=0 400 450 0,clip, scale = 0.44]{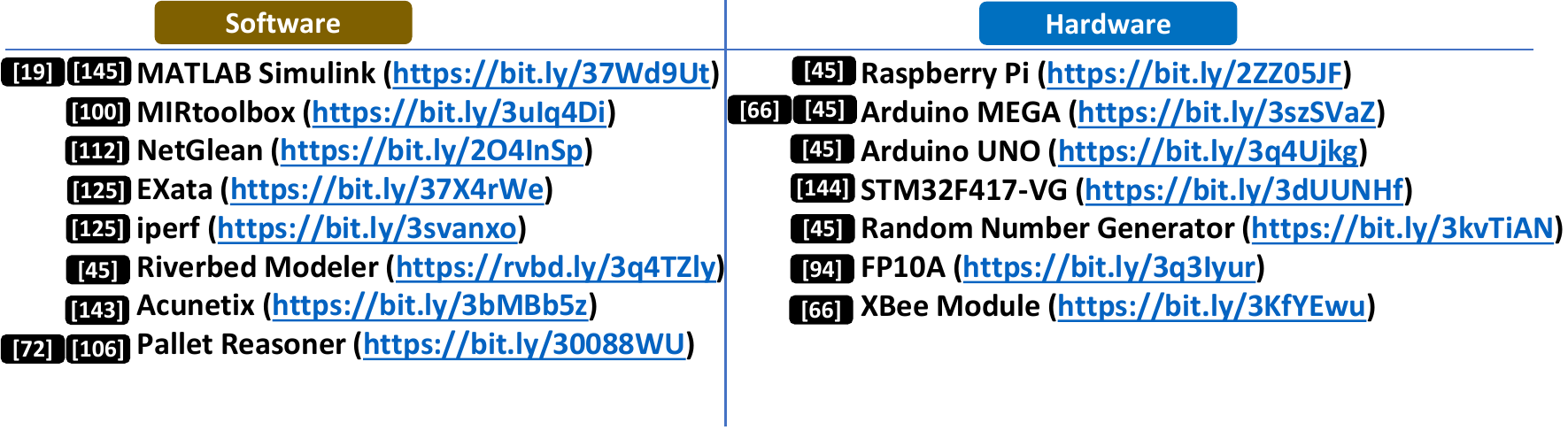}
  \caption{Software and hardware tools used for C3I systems security evaluation}
  \label{fig:tools}
\end{minipage}
\end{figure}

\subsubsection{Need of rigorous and comprehensive evaluation}
We investigated the evaluation methods applied in the surveyed articles and identified that 36.67\% studies have not evaluated their proposed security mechanisms (Figure~\ref{fig:evaluation}). Moreover, 25\% of the studies have been evaluated through hypothetical scenarios only. Accordingly, it is evident that over 61\% of research have not followed a systematic and rigorous evaluation process in their studies. From the surveyed papers, 5\%, 16.67\% and 15\% studies have been evaluated via prototyping, simulation, and laboratory experiments, respectively. Figure~\ref{fig:tools} presents the hardware and software tools used in these evaluation methods. Notably, only one study has been evaluated in a real-world setting. Therefore, it raises significant concerns on the reliability of the existing literature due to the lack of rigorous evaluation. Furthermore, it is being exacerbated by the highly technical and critical nature of C3I domain. As discussed in Section \ref{overview}, C3I systems consist of multiple heterogeneous subsystems, and some of these are deployed in hostile environments. Accordingly, practical issues such as physical destruction of nodes due to hostile conditions, compromised software components, degradation of hardware components with time, and natural disasters (e.g., flood, fire, storms) present a significant impact on the performance, security, and safety of C3I systems. Therefore, we advocate that countermeasures for protecting C3I systems should be evaluated in real-world settings through a systematic evaluation approach to strengthen the credibility of the resulting research finding.

\section{Conclusion} \label{conclusion}

The cybersecurity of C3I systems has become a serious concern due to the severe criticality of tactical domains (e.g., military and rescue missions) where C3I systems are used. Therefore, we gather, examine, and synthesize literature on the cybersecurity of C3I systems. Based on our review, we have critically analyzed and categorized 13 security vulnerabilities, 19 attack vectors, and 40 countermeasures considered important for the cybersecurity of C3I systems. Moreover, we have presented our analysis on the research findings, which includes: (i) interrelation of attack vectors and security vulnerabilities; (ii) interrelation of attack vectors and countermeasures; and (iii) identification of future research directions for advancing the field of C3I system's security.

This survey provides several benefits for both researchers and practitioners. In particular, for researchers, our survey has proposed several future research directions for further exploring and strengthening the cybersecurity of C3I systems. For example, the SOA adaptation for C3I systems introduces many challenges (e.g., lack of continuous monitoring) in securing the cyberspace of C3I systems. Similarly, the incorporation of cloud computing into C3I systems raises numerous security concerns for sensitive data confidentiality and integrity. For practitioners, the mapping of attack vectors to countermeasures enable C3I systems' operators to identify an accurate countermeasure against an attack vector. Similarly, the need of prioritizing countermeasures while designing C3I systems helps system engineers in implementing optimal countermeasures during the development of C3I systems. We hope that the findings reported in this survey will provide researchers and practitioners new dimensions and inspirations for pushing their research and development efforts to secure C3I systems.

\begin{acks}
The work has been supported by the Cyber Security Research Centre Limited whose activities are partially funded by the Australian Government’s Cooperative Research Centres Programme.
\end{acks}


% \nociteS{*}
% \bibliographystyleS{unsrt}
% \bibliographyS{appendix}

\bibliographystyle{ACM-Reference-Format}
\bibliography{acmart}

\end{document}